\documentclass[11pt,a4paper]{article}
\usepackage{jcappub}

\usepackage{bm}
\usepackage{epsfig}
\usepackage{citesort}
\usepackage{graphicx}
\usepackage{amssymb}
\usepackage{color}
\usepackage{subfigure}

\begin{document}


\title{Diffuse emission of high-energy neutrinos from gamma-ray burst fireballs}

\author[a]{Irene Tamborra}
\author[a]{and Shin'ichiro Ando}

\affiliation[a]{GRAPPA Institute, University of Amsterdam, Science Park 904, 1098 XH, Amsterdam, The Netherlands}

\emailAdd{i.tamborra@uva.nl}
\emailAdd{s.ando@uva.nl}

\abstract{
Gamma-ray bursts (GRBs)  have been suggested as possible sources of the
high-energy neutrino flux recently detected by the IceCube telescope.
We revisit the fireball emission model and elaborate an
analytical prescription to estimate the high-energy neutrino prompt
emission from pion and kaon decays, assuming that the leading mechanism
for the neutrino production is lepto-hadronic. To this purpose, we
include  hadronic, radiative and adiabatic cooling effects and discuss
their relevance for long- (including high- and low-luminosity) and
short-duration GRBs.
The expected diffuse neutrino background is derived, by requiring that
the GRB high-energy neutrino counterparts follow up-to-date gamma-ray
luminosity functions and redshift evolutions of the long and short GRBs.
Although dedicated stacking searches have been unsuccessful up to now,
we find that  GRBs could contribute up to a few $\%$ to the observed
IceCube high-energy neutrino flux for sub-PeV energies, assuming that
the latter has a diffuse origin.  Gamma-ray bursts, especially low-luminosity ones, 
could however be the main sources of the IceCube high-energy neutrino flux  in the PeV range.
 While high-luminosity  and low-luminosity GRBs have comparable intensities, 
  the contribution from the short-duration component is significantly
smaller.
Our findings confirm the most-recent IceCube results on the GRB searches
and suggest that larger exposure is mandatory to detect high-energy
neutrinos from  high-luminosity GRBs  in the near future.
}
\maketitle

\section{Introduction}         \label{intro}  
Gamma-ray bursts (GRBs) are among the most energetic events in the
Universe  (see e.g.~\cite{Kumar:2014upa,Meszaros:2006rc} for reviews on
the topic) and have been suggested as  sources of ultra-high energy
cosmic rays~\cite{Waxman:1995vg,Vietri:1995hs}. In terms of astronomical
observations, they  are usually divided into two distinct groups on the
basis of the BATSE bimodal distribution: Long-duration bursts (whose
duration is longer than $2$~s) and short-duration bursts (lasting for
less than $2$~s)~\cite{Kouveliotou:1993yx}. Long duration bursts  are
thought to  originate from the collapse of a massive star to a black
hole~\cite{Woosley:1993wj}, while short-duration ones should originate
from coalescing neutron stars or black-hole--neutron-star
mergings~\cite{Eichler:1989ve}.  

It is not excluded that, although with different origin, short- and
long-duration bursts are driven by the same underlying mechanism: The
fireball model~\cite{Waxman:1997ti,Waxman:2003vh,Hummer:2011ms}. 
According to this model, a hot ``fireball'' of electrons, protons and
photons forms. Such fireball  is initially opaque to radiation, then the
hot plasma expands by radiation pressure and particles are accelerated
making  the plasma  transparent to radiation and favoring the emission
of  keV-MeV photons~\cite{Shemi:1990rv,Rees:1992ek}.  Neutrinos with
energies of $\mathcal{O}(100)$~TeV are also expected to be emitted from
these sources because of lepto-hadronic interactions~\cite{Waxman:1997ti}.
Besides the fireball emission model~\cite{Waxman:1997ti,Hummer:2011ms}
on which we will relay in this work, other models have been proposed to
explain the neutrino production in GRBs, such as the dissipative
photosphere model~\cite{Rees:2004gt} 
according to which the prompt GRB emission occurs near the Thomson
scattering photosphere, or large-radius magnetic dissipation models,
such as the ICMART model that relies on a highly magnetised outflow
which is dissipated at a radius larger than the internal
shock radius~\cite{Zhang:2010jt}. See~\cite{Zhang:2012qy} for a common
formalism to connect these models and a comparison among the expected
neutrino fluxes.

The IceCube detector, a neutrino telescope made with 5160 optical
modules and located at the South Pole, could detect neutrinos from GRBs
by measuring the Cherenkov light from secondary  particles produced in
the neutrino-nucleon interactions.
Over the past years, IceCube performed searches for  muon neutrinos
associated to GRBs,  but with negative
results~\cite{Abbasi:2009ig,Abbasi:2011qc,Abbasi:2012zw,Aartsen:2013dla,Aartsen:2014aqy}.
The IceCube upper limits therefore started to put tight constraints on
the expected GRB flux and on the  theoretical models employed  to
explain the neutrino emission from these
sources~\cite{Baerwald:2011ee,He:2012tq}.
The diffuse neutrino emission from GRBs is also widely discussed in the
literature~\cite{Waxman:1998yy,Murase:2005hy,Murase:2006mm,Gupta:2006jm,Liu:2011cua,Asano:2014nba,Ahlers:2011jj,Baerwald:2014zga,Bustamante:2014oka}
and it has been recently
invoked~\cite{Liu:2012pf,Cholis:2012kq,Murase:2013ffa,Wang:2015xpa,Razzaque:2014ola,Razzaque:2013dsa,Petropoulou:2014lja,Nakar:2015tma}
as a natural possibility to explain the PeV neutrino events discovered
from IceCube~\cite{Aartsen:2013bka,Aartsen:2013jdh,Aartsen:2014gkd,Aartsen:2014muf},
assuming such events have a diffuse origin (see
\cite{Anchordoqui:2013dnh} for an overview on the possible astrophysical
sources of the IceCube PeV neutrinos and references therein).

In light of the unsuccessful most recent IceCube  stacking
analysis~\cite{Aartsen:2014muf} and of the high-energy neutrino flux
discovery~\cite{Aartsen:2013bka,Aartsen:2013jdh,Aartsen:2014gkd,Aartsen:2014muf},
the aim of this work is to provide an up-to-date estimation of the
expected high-energy neutrino prompt emission from long- and
short-duration GRB families within the fireball
model~\cite{Waxman:2003vh,Hummer:2011ms}.
To this purpose, for the first time we present an analytically extensive
modelling of the neutrino emission from fireballs by including
radiative, adiabatic and hadronic cooling processes involving pion and
kaon decays.
Then, we discuss the relevant processes for each GRB family
(high-luminosity, low-luminosity and short GRBs) and derive  an
estimation of the diffuse flux by  adopting up-to-date gamma-ray
luminosity functions and evolution of formation rates to fix the
normalization of the GRB neutrino energy spectrum.

Our results show that, for average GRB parameters compatible with
observations, the  diffuse neutrino emission from GRBs  could contribute
up to a few $\%$ to the currently observed IceCube high-energy neutrino
flux  in the sub-PeV region while GRBs might be the main source
of the IceCube flux in the PeV range, assuming that the latter has a  diffuse origin.
Moreover, we find that the neutrino emission within the fireball model
is widely compatible with the current IceCube limits due to stacked
GRBs.

This manuscript is organised as follows. In
Section~\ref{sec:observations} we introduce constraints coming from
gamma-ray observations on the luminosity functions and redshift
distribution of long-duration (divided in low-luminosity and
high-luminosity GRBs) and short-duration GRBs. In
Sec.~\ref{sec:cooling}, we analytically model the neutrino emission from
$p\gamma$ interactions in GRBs and define the expected neutrino
spectrum from pion and kaon decays including  radiative, adiabatic and
hadronic cooling processes. In Sec.~\ref{sec:nuback}, we present our
results on the diffuse emission from these sources,  discussing the 
astrophysical uncertainties as well as the ones related to the GRB model parameters,
and compare them with  IceCube bounds and future searches. Conclusions are
presented in Sec.~\ref{sec:conclusions}.

\section{Observational constraints on gamma-ray bursts}\label{sec:observations}
Gamma-ray bursts have been monitored since long time now. They are
usually divided into two families according to the observation of these
transients in photons: Long- and short-duration bursts.  However,
current data are still insufficient to define accurate luminosity
functions, especially at high redshifts. Sources of errors are, for
example, the  triggering criteria of the detection instruments that can
be responsible for poor estimates in the parameters,  degeneracies
arising from a mixing of the luminosity function and the source rate
evolution with the redshift, as well as  selection effects. Given such
uncertainties, in the following we will define  allowed bands for the
distribution of the GRB families in luminosity and redshift and
characterize     the source energetics. 
Unless otherwise specified, we will distinguish among three different
reference frames: The GRB reference frame, the jet comoving frame and
the observer frame. Each physical quantity $X$ will be labelled as
$\tilde{X}$, $X^{\prime}$, and $X$ in each of these frames
respectively.

\begin{table}
  \begin{center}
  \caption{GRB parameters adopted in the estimation of the diffuse high-energy neutrino flux  for our \emph{canonical} model including the astrophysical uncertainties. The local rate ($\rho_0$) is in units of Gpc$^{-3}$~yr$^{-1}$, the isotropic luminosities ($\tilde{L}_\star$, $\tilde{L}_{\rm min}$ and $\tilde{L}_{\rm max}$) are expressed in units of $10^{52}$~erg s$^{-1}$, the variability time $t_v$ is in s.
  The best fit parameters $\alpha$ and $\beta$ are employed in the luminosity function fits, while $\alpha_\gamma$ and $\beta_\gamma$ describe the gamma-ray Band spectrum, and $\Gamma$ is the bulk Lorenz factor of the jet.}
  \begin{tabular}{rrrrrrrrrrr}\hline\hline
   \  & $\rho_0$ & $\tilde{L}_{\star}$  & $\alpha$ & $\beta$ &$\tilde{L}_{\rm min}$ &$\tilde{L}_{\rm max}$ & $\alpha_\gamma$ & $\beta_\gamma$ & $\Gamma$ & $t_v$ \\ \hline 
   HL-GRB & $0.8$ & $0.8$ & $-0.95$ & $-2.59$ & $10^{-3}$ & $10^2$ & $1$ & $2$ & $500$ & $0.1$\\
   HL-GRB & $0.5$ & $0.48$ & $-0.13$ & $-2.42$ & $10^{-3}$ & $10^2$ & $1$ & $2$ & $500$ & $0.1$\\
   \hline
  LL-GRB & $2000$ & $5 \times 10^{-4}$ & $-2.3$ & $-1.27$ & $1.8\times 10^{-6}$ & $10^{-3}$ & $1$ & $2$ & $5$ & $100$\\
 LL-GRB & $200$ & $5 \times 10^{-4}$ & $-2.3$ & $-1.27$ & $1.8\times 10^{-6}$ & $10^{-3}$ & $1$ & $2$& $5$ & $100$\\
 \hline
   sGRB & $4.6_{-1.7}^{+1.9}$ & $2$ & $-1.94$ & $-3.0$& $0.5\times 10^{-3}$ & $10$ & $0.5$ & $2.25$ & $650$ & $0.01$\\
    \hline\hline \label{table:LF}
  \end{tabular}
 \end{center}
\end{table}

\subsection{Long duration gamma-ray bursts}

Long-duration GRBs are thought to originate from the collapse of a
massive star into a black hole~\cite{Woosley:1993wj}.
They are usually divided in two sub-categories: High-luminosity (HL) and
low-luminosity (LL) GRBs.

The source rate evolution as a function of the redshift ($z$) of the HL
component is described through a piecewise
function~\cite{Wanderman:2009es}:
\begin{eqnarray}
\label{eq:rateHLGRB}
 R_{\rm HL-GRB} = \rho_0 \left\{ \begin{array}{lll}
 (1+z)^a & \mathrm{for} & z < z_{\star} \\ 
 (1+z_\star)^{a-b} (1+z)^b & \mathrm{for} & z \ge z_{\star}\ ,
\end{array}\right.
\end{eqnarray} 
being $\rho_0$ the local rate defined as in Table~\ref{table:LF},
$z_\star = 3.6$, $a = 2.1$, and $b = -0.7$~\cite{Wei:2013wza}.  

\begin{figure}[!t]
\begin{center}  
\includegraphics[width=0.95\columnwidth]{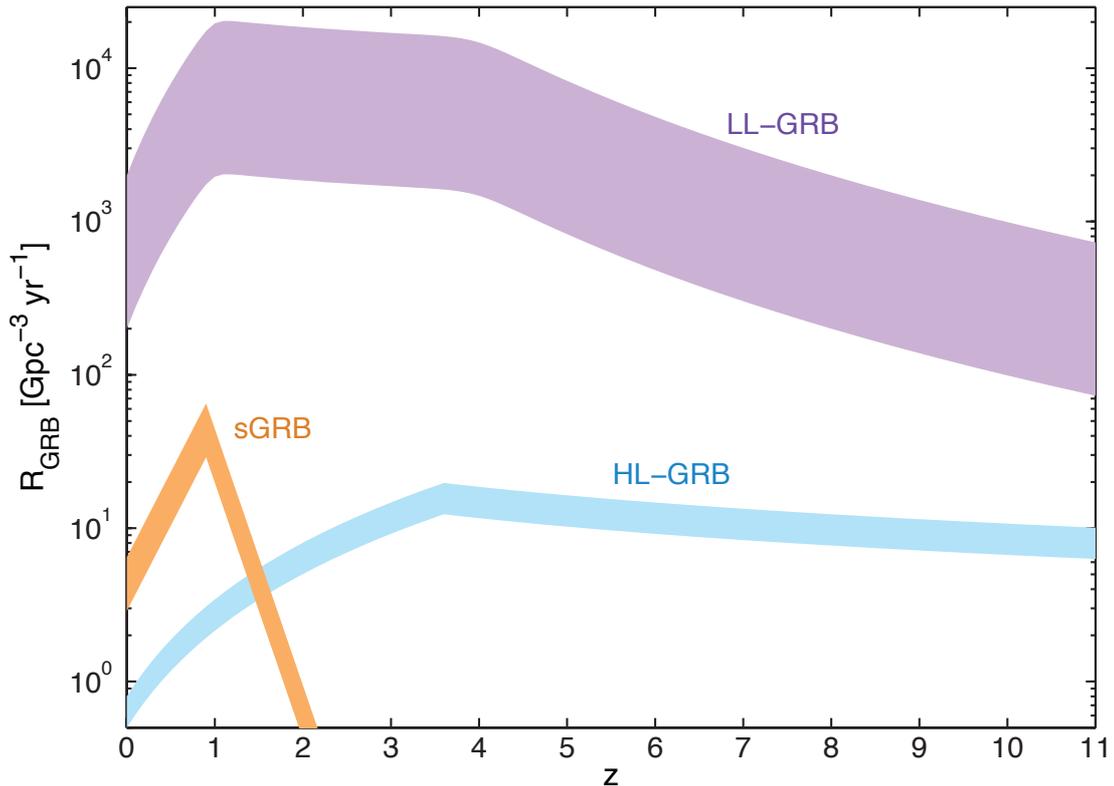}  
\end{center}  
\caption{Redshift distribution of the HL-GRBs (light blue),  LL-GRBs (violet), and sGRBs (orange). Each family is normalised to its local rate $\rho_0$. The LL-GRB local rate is higher than the HL-GRB and 
 the sGRB ones, and the sGRB rate quickly decreases for  $z \ge 1$. \label{fig:rate}}  
\end{figure}

The HL-GRB typical luminosities vary in the range
$[10^{49},10^{54}]$~erg~s$^{-1}$. 
We adopt the intrinsic isotropic luminosity function (LF), i.e.,
corrected for beaming effects and defined in the GRB frame, found in
\cite{Howell:2014wba} on the basis of a selected sample of 175 bursts
observed by {\it Swift}:
\begin{eqnarray}
\label{eq:ratesGRB}
 \Phi_{\rm HL-GRB} \propto  \left\{ \begin{array}{lll}
 \left(\frac{\tilde{L}_{{\rm iso}}}{\tilde{L}_{\star}}\right)^{\alpha} &
  \mathrm{for} & \tilde{L}_{{\rm iso}} < \tilde{L}_{{\star}} \\ 
  \left(\frac{\tilde{L}_{{\rm iso}}}{\tilde{L}_{\star}}\right)^{\beta} &
   \mathrm{for} & \tilde{L}_{{\rm iso}} \ge \tilde{L}_{{\star}}\ ,
\end{array}\right.
\end{eqnarray} 
with best-fit parameters defined in
Table~\ref{table:LF}~\cite{Howell:2014wba}. Note that in
Table~\ref{table:LF} we consider two sets for the LF parameters and for
the local rate $\rho_0$, corresponding to the extremes of the accepted
band allowed from the data (see~\cite{Howell:2014wba} for more details).
Figure~\ref{fig:rate} shows the redshift distribution of the HL-GRB
family in light blue.  The uncertainty band is defined by the two extreme sets of parameters in Table~\ref{table:LF}
for HL-GRBs.  The distribution of these sources peaks at about
$z \simeq z_\star$ and it stays roughly constant at higher redshifts.

The LL-GRBs are characterised by isotropic equivalent luminosity much
smaller than that of the HL-GRBs,
i.e. $10^{46}$--$10^{49}~\mathrm{erg~s}^{-1}$. 
The origin of the LL component is not yet clear.
It is  not excluded  that the LL-GRBs   belong to a distinct
population than the HL-GRBs~\cite{Liang:2006ci}. 
On the other hand, it could be that they have the same origin as
HL-GRBs, but are produced by failed jets that do not break out of their
progenitors.
According to the latter hypothesis, this could mean that most supernovae
generate jets  producing LL-GRBs and only few of them are able to
produce  jets powerful enough to  break out and produce
HL-GRBs~\cite{Bromberg:2011fm}. 
Due to their low-luminosity, LL-GRBs  have been mainly detected nearby
($z \le 0.1$) and a large fraction of their population might be below
the detection threshold.

In order to estimate the  LL-GRB total event rate, we assume a
supernova--GRB connection, and suppose that the LL-GRB rate should be
lower than the one of type Ib/c supernovae.
Therefore, we define the LL-GRB rate as~\cite{Liu:2011cua,Yuksel:2008cu}:
\begin{eqnarray}
       R_{\rm LL-GRB}=\rho_0\left[(1 + z)^{p_1 \kappa} + \left(\frac{1 + z}{5000}\right)^{p_2 \kappa} +
        \left(\frac{1 + z}{9}\right)^{p_3 \kappa}\right]^{1/\kappa}\ ,
\end{eqnarray}
with $\kappa = -10$, $p_1 = 3.4$, $p_2 = -0.3$, $p_3 = -3.5$, and we
consider two representative values for the fraction of SN Ib/c that goes
in LL-GRBs~\cite{Liu:2011cua}, i.e., $\rho_0 = 0.01\ \rho_{\rm SN}$ and
$\rho_0 = 0.1\ \rho_{\rm SN}$ with $\rho_{\rm SN} = 2 \times
10^4$~Gpc$^{-3}$~yr$^{-1}$~\cite{Dahlen:2004km}. The LL-GRB rate is
shown as a function of the redshift in Fig.~\ref{fig:rate} (violet
band), it peaks at $z \simeq 1$ and it decreases at higher
redshifts. Note that it is locally much higher than the HL-GRB one.

The LL-GRB LF is parametrized as in~\cite{Liu:2011cua,Dai:2008mw}:  
\begin{equation}
\Phi_{\rm LL-GRB} \propto \left[\left(\frac{\tilde{L}_{\rm iso}}{\tilde{L}_\star}\right)^\alpha + \left(\frac{\tilde{L}_{\rm iso}}{\tilde{L}_\star}\right)^\beta\right]\ ,
\end{equation}
 with  $\tilde{L}_\star$, $\alpha$ and $\beta$ as in Table~\ref{table:LF} based on  the {\it Swift}-BAT  sample analysed in~\cite{Dai:2008mw}.   
Although the adopted LF fit  describes both the LL and
HL-GRBs~\cite{Dai:2008mw}, we use it  for the LL-GRB component only.

For both the LL- and HL-GRBs, we assume that the  injected (inj) gamma-ray energy
spectrum is  fitted with a Band-spectrum~\cite{Band:1993eg}:
\begin{eqnarray}
\label{eq:gammaspectrum}
\left(\frac{dN_\gamma}{dE^\prime_{\gamma}}\right)_{\rm inj} \propto  \left\{ \begin{array}{lll}
  \left(\frac{E^{\prime}_{\gamma,b}}{E^\prime_\gamma}\right)^{\alpha_\gamma}
   & \mathrm{for} & E^\prime_{\gamma} < E^{\prime}_{\gamma,b}\ ,\\ 
  \left(\frac{E^{\prime}_{\gamma,b}}{E^\prime_{\gamma}}\right)^{\beta_\gamma}
   & \mathrm{for} & E^\prime_{\gamma} \ge E^{\prime}_{\gamma,b}\ ,
\end{array}\right .
\end{eqnarray}
with $\alpha_\gamma = 1$ and $\beta_\gamma = 2$ as in Table~\ref{table:LF}.
The photon break energy $E_{\gamma,b}$  is usually expressed as a
function of the isotropic energy ($E_{\rm iso}$) through the so-called
``Amati relation'' that holds for the long-duration
GRBs~\cite{Amati:2002ny,Ghirlanda:2005rq}: 
\begin{equation}
\label{eq:EbEiso}
\frac{\tilde{E}_{\gamma,b}}{0.1\ {\rm MeV}}  =  (3.64 \pm 0.04) \left(\frac{\tilde{E}_{{\rm iso}}}{7.9 \times 10^{52}\ \mathrm{erg}}\right)^{0.51\pm 0.01} \ .
\end{equation}

The isotropic energy of each GRB in the source rest frame can be
expressed as a function of the isotropic luminosity ($\tilde L_{\rm
iso}$) by combining the  ``Yonetoku relation''~\cite{Yonetoku:2003gi}
and the Amati one~\cite{Ghirlanda:2011bn} for the long-duration
GRBs~\cite{Liu:2012pf}:
\begin{equation}
\label{eq:LisoEiso}
\log\left(\frac{\tilde{E}_{{\rm iso}}}{10^{52}\ \mathrm{erg}}\right) = 1.07 \log\left(\frac{\tilde{L}_{{\rm iso}}}{10^{52}\ \mathrm{erg}/\mathrm{s}}\right) + (0.66 \pm 0.54)\ .
\end{equation}
In the following, we will assume that the Amati and Yonetoku relations
hold for both the LL- and HL-GRBs, postulating that the two populations
share the same emission mechanism (see
e.g.~\cite{Liang:2006ci,Guetta:2006gq} for a discussion on the topic). 

The photon break energy in the jet frame is related to the same quantity
in the GRB frame through $E^{\prime}_{\gamma,b} =
\tilde{E}_{\gamma,b}/\Gamma$. 
The normalisation of the photon spectrum, $f_\gamma$, is set by assuming
that $f_\gamma = \tilde E_{{\rm iso}}/\int_{0}^{\infty} d\tilde
E_{\gamma} \tilde E_{\gamma} (dN_\gamma/d\tilde E_{\gamma})$.

\subsection{Short-duration gamma-ray bursts}

Short-duration GRBs (sGRB) exhibit typical luminosities similar to those
of the HL-GRBs, and are believed to originate from neutron-star--neutron-star or
neutron-star--black-hole mergers~\cite{Eichler:1989ve}. Therefore, we
should expect a delay between the star-formation rate and the merger
rate due to the spiral-in time~\cite{Narayan:1992iy,Ando:2004pc}.
In order to model the sGRB redshift distribution and their luminosity
function, we follow~\cite{Wanderman:2014eza} which selects a sample of
non-collapsar GRBs from the BATSE, {\it Swift}, and {\it Fermi} data.

The sGRB rate is described by the convolution of the ordinary star-formation rate with a
function $f(\Delta t)$ that takes into account the time-delay $\Delta t$
of the spiral-in time of the
binaries~\cite{Narayan:1992iy,Ando:2004pc}.
According to~\cite{Wanderman:2014eza},  the time-delay function that is
in better agreement with the data  is a log-normal distribution that
allows to express the sGRB formation with the following effective rate:
\begin{equation}
R_{\rm sGRB} = 10\ \rho_0\ \left\{ \begin{array}{lll}
 \exp[(z - 0.9)/0.39]& \mathrm{for}& z \le z_\star\\ 
  \exp[-(z - 0.9)/0.26]& \mathrm{for}& z > z_\star\ ,
  \end{array}\right.
\end{equation}
with $z_\star = 0.9$.
The sGRB rate as a function of the redshift is plotted in
Fig.~\ref{fig:rate} (orange band), note as it peaks at about $z_\star$
and then it decreases quickly. The sGRB rate  is locally higher than the
HL-GRB one.

The LF is fitted with a broken power law~\cite{Wanderman:2014eza}:
\begin{eqnarray}
\label{eq:ratesGRB}
 \Phi_{\rm sGRB} \propto  \left\{ \begin{array}{ll}
 \left(\frac{\tilde{L}_{\rm iso}}{\tilde{L}_\star}\right)^{\alpha}& \mathrm{for}\ \tilde{L}_{\rm iso} \le L_{\star} \\ 
  \left(\frac{\tilde{L}_{\rm iso}}{L_\star}\right)^{\beta}& \mathrm{for}\ \tilde{L}_{\rm iso} > L_{\star}\ ,
\end{array}\right.
\end{eqnarray} 
with the best fit parameters provided in Table~\ref{table:LF}.

Similarly to the long-duration GRBs, sGRBs have a gamma-ray spectrum
fitted with the Band spectrum (Eq.~\ref{eq:gammaspectrum}). However, we
know from observations that the low-energy component (i.e., for $E_\gamma < E_{\gamma,b}$) is harder for sGRBs
 than for the long-duration GRBs (see the values for $\alpha_\gamma$ in Table~\ref{table:LF}) and the peak
energy is slightly higher~\cite{Nava:2010ig,Wanderman:2014eza,Shahmoradi:2014ira}.

We assume that relations similar to the Amati  and Yonetoku  ones hold
between $\tilde{E}_{\gamma,b}$, $\tilde{E}_{\rm iso}$, and
$\tilde{L}_{\rm iso}$ for the sGRBs. To this purpose,
we extrapolate them by fitting the data in Fig.~7 of \cite{Shahmoradi:2014ira} and define the analogous of Eqs.~(\ref{eq:EbEiso}) and (\ref{eq:LisoEiso}):
\begin{equation}
\log\left(\frac{\tilde{E}_{\gamma,b}}{{0.1\ {\rm MeV}}}\right) = 0.56
 \log\left(\frac{\tilde{E}_{{\rm iso}}}{10^{52}\ \mathrm{erg}}\right) +
 3.23\ , 
\end{equation}
\begin{equation}
\log \left(\frac{\tilde{E}_{{\rm iso}}}{10^{52}\ \mathrm{erg}}\right)
 = 1.06 \log\left(\frac{\tilde{L}_{{\rm iso}}}{10^{52}\
	       \mathrm{erg}/\mathrm{s}}\right) - 1.57\ .
\end{equation}
We suppose that this class of GRBs has shorter variability timescale ($t_v$) than long-duration GRBs~\cite{MacLachlan:2012cd}, as reported in Table~\ref{table:LF}.

\section{Prompt neutrino emission  from gamma-ray burst fireballs}\label{sec:cooling}
In this Section, we discuss the neutrino production in  GRBs through
$p\gamma$ interactions and derive the corresponding neutrino energy
distributions. The main reactions that we study  are:
\begin{eqnarray}
p + \gamma &\rightarrow&  \Delta \rightarrow n + \pi^{+}, p + \pi^{0}\  \\ 
\ p + \gamma &\rightarrow& K^+ + \Lambda/\Sigma\ . \nonumber
\end{eqnarray}
Pions, kaons and neutrons in turn decay into neutrinos:
\begin{eqnarray}
&\pi^{+}& \rightarrow \mu^+ \nu_\mu\ ,\\ \nonumber
&\mu^+& \rightarrow \bar{\nu}_\mu + \nu_e + e^+\ ,\\ \nonumber
&\pi^{-}& \rightarrow \mu^- \bar{\nu}_\mu\ ,\\ \nonumber
&\mu^-& \rightarrow \nu_\mu + \bar{\nu}_e + e^-\ ,\\ \nonumber
&K^+& \rightarrow \mu^+ + \nu_\mu\ ,\\ \nonumber
&n& \rightarrow p + e^- + \bar{\nu}_e\ . \nonumber
\end{eqnarray}
In the following, we will assume that the neutrino contribution from the $n$ decay is negligible (see Fig.~2 of \cite{Baerwald:2010fk}) and we will reconstruct the neutrino energy spectrum from the pion and kaon decays.

\subsection{Neutrino production from pion decay} 
The comoving proton  energy threshold to produce a $\Delta$ resonance is  
$E^{\prime}_{p} \ge  [(m_\Delta c^2)^2 - (m_p c^2)^2]/(4
E^\prime_{\gamma})$, where  $m_\Delta$ ($m_p$) is the $\Delta$ ($p$)
mass. 
The corresponding neutrino energy in the observer frame
is~\cite{Guetta:2003wi}:
$E_{\nu} = 0.05\ [\Gamma/(1+z)]^2\ (m_\Delta^2 - m_p^2)/(2 E_{\gamma})$,
with  $\Gamma$ the bulk Lorenz factor of the jet that we assume constant
for sake of simplicity; the numerical factor comes from the fact that
the average energy fraction transferred from the initial proton to the
pion is $20\%$ times $1/4$ which comes from the assumption that the $4$
leptons in the $\pi$ decay channel  equally share the energy of the pion
(i.e., $E_{\nu} = 0.05\ E_{p}$).

The resultant neutrino spectrum will exhibit a spectral break at:
\begin{equation}
E_{\nu,b,\pi} = a_\pi \left(\frac{\Gamma}{1+z}\right)^2\ \frac{(m_\Delta c^2)^2 - (m_p c^2)^2}{2 E_{\gamma, b}} = 1.6 \times 10^6 E_{\gamma, b,{\mathrm{MeV}}}^{-1} \left(\frac{\Gamma_{2.5}}{1+z}\right)^2\ \mathrm{GeV}\ ,
\label{eq:firstbreak}
\end{equation}
where, in the first equality, the correction factor of 2 with respect to $E^\prime_{p}$ takes
into account the fact that the pion production efficiency peaks at
higher center of mass energies~\cite{Baerwald:2010fk}. The pre-factor
$a_\pi$ for  neutrinos produced by pion decay is $a_{\pi} = 0.05 = 0.2
\times 1/4$ (since $20\%$ is the fraction of the proton energy that goes
into pion and $1/4$ is the fraction of the $\pi$ energy carried by
neutrinos).  In order to favor a comparison with the
existing literature, in the second equality, we express it in terms of
typical values for the GRB parameters: $\Gamma_{2.5} = \Gamma/10^{2.5}$ and 
$E_{\gamma, b,{\mathrm{MeV}}} = E_{\gamma, b}/\mathrm{MeV}$.

Above $E_{\nu,b,\pi}$, the neutrino spectrum is the same
as the proton spectrum, because the protons can interact with photons with energies $E_{\gamma,b}$ or smaller, where
 the number of photons is almost constant,
i.e., $E_\gamma dN_\gamma / dE_\gamma \propto E_\gamma^{1-\alpha_\gamma}
= E_\gamma^0$. Below $E_{\nu, b, \pi}$, on the other hand, the
number of photons  with energy $E_\gamma$ that a proton with energy
$E_\nu/a_\pi$ can interact with is suppressed by a factor of
$(E_\gamma/E_{\gamma, b})^{1-\beta_\gamma} = (E_{\nu, b, \pi} /
E_\nu)^{1-\beta_\gamma}$, and the resulting neutrino spectrum is harder
by a factor of $E_\nu^{\beta_\gamma-1}$  with respect to the proton
spectrum (that we assume to scale as $\sim E_p^{-2}$).

If pions produced by the $p\gamma$ interactions decay faster than they
cool, the correspondent neutrino spectrum is not affected.
Otherwise, other neutrino break energies  are
determined by the radiative cooling (rc; synchrotron radiation and
inverse Compton scattering), adiabatic cooling (ac), and hadronic
cooling (hc) processes, being the total cooling timescale  defined as
$t^{\prime -1}_c = t^{\prime -1}_{\rm hc} + t^{\prime -1}_{\rm rc} +
t^{\prime -1}_{\rm ac}$.
The neutrino spectrum is modified by an additional factor
of $[1-\exp(-t_c^\prime m_\pi / E_\pi^\prime \tau_\pi)]$, where $\tau_\pi$
is the pion lifetime and $m_\pi$ is the pion mass.
If the cooling time scale is much shorter than the decay lifetime in the jet
frame (i.e., $t_c^\prime \ll E_\pi^\prime \tau_\pi / m_\pi$), the additional factor can be
approximated as $t_c^\prime m_\pi / E_\pi^\prime \tau_\pi$.
Below, we shall discuss the three relevant cooling processes.

The hadronic cooling  timescale for pions in the jet frame is: 
\begin{equation}
\label{eq:tprimehc}
t^{\prime}_{\pi,{\rm hc}} = \frac{E^{\prime}_{\pi}}{c \sigma_h n^{\prime}_{p} \Delta E^{\prime}_{\pi}} =  \frac{20 \pi c^4 \Gamma^6 m_p t_{v}^2 \epsilon_e}{0.3 \sigma_h \tilde{L}_{\rm iso} (1+z)^2}\ ,
\end{equation}
where $c$ is the speed of light, $n^\prime_{p} = E_{j} (1+z)/(\Gamma m_p
c^2 V^\prime)$ the jet comoving proton density, $\sigma_h = 5 \times
10^{-26}$~cm$^{-2}$~\cite{Eidelman:2004wy} the cross section for
meson-proton collisions, and $\Delta E^{\prime}_\pi$  the energy lost by
the incident meson in each collision, which is $\Delta E^\prime_\pi =
0.8 E^\prime_\pi$~\cite{Brenner82}. For a jet with kinetic energy $E_j$,
opening angle $\theta_j$ and variability timescale $t_{v}$, the volume
of an infinitesimal jet shell is $V^\prime = 2 \pi \theta_j^2
\tilde{r}_j^2 c t_j \Gamma/(1+z)$ with   $\tilde{r}_j = 2 \Gamma^2 c
t_v/(1+z)$ the internal shock radius.
We then converted the jet energy $E_j$ divided by the duration $t_j$
(the jet luminosity $L_j = E_j / t_j$) into the isotropic peak
luminosity via $L_j/(2\pi\theta_j^2) = 0.3 \tilde{L}_{\rm
iso}/[4\pi(1+z)^2 \epsilon_e]$, where $0.3$ comes from the fact that typically 0.3
times the peak luminosity is the luminosity averaged over the duration
of the burst~\cite{Kakuwa:2011aq, Liu:2012pf},  and  $\epsilon_e$
is the energy fraction carried by the electrons.
When the hadronic cooling is the dominant cooling process (i.e.,
$t^\prime_c \simeq t^\prime_{\pi,{\rm hc}}$), a steepening of the
neutrino energy spectrum occurs at an energy satisfying $t^\prime_{\rm hc}
= \tau_\pi E^\prime_\pi/(m_\pi c^2)$ with $\tau_\pi$ ($m_\pi$) the pion
lifetime (mass), i.e.,
\begin{equation}
E_{\nu,\pi,{\rm hc}} =  \frac{b_\pi \Gamma}{(1+z)^3}\ \frac{20 \pi m_\pi m_p c^6 \Gamma^6 t_v^2 \epsilon_e}{0.3 \tau_\pi \sigma_h \tilde{L}_{\rm iso}}  = 1.75 \times 10^{13} \frac{\Gamma_{2.5}^7 t_{v,-2}^2 \epsilon_e}{(1+z)^3 \tilde{L}_{{\rm iso},52}} \ \mathrm{GeV}\ ,
\label{eq:hcpi}
 \end{equation} 
where $b_\pi = 1/4$, $E^\prime_\nu = E^\prime_\pi/4$, $t_{v,-2} = t_v /
(10^{-2}~{\rm s})$, and $\tilde L_{\rm iso, 52} = \tilde L_{\rm iso} /
(10^{52}~{\rm erg~s}^{-1})$.
Since $t^{\prime}_{\pi,{\rm hc}}$ is independent of
energy, the neutrino spectrum suppression factor due to the hadronic
cooling is $t_{\rm hc}^\prime m_\pi / (E_\pi^\prime \tau_\pi) \propto
1/E^{\prime}_{\nu}$.

The adiabatic cooling time in the jet comoving frame is given by
\begin{equation}
 \label{eq:tprimeac}
t^\prime_{\pi,{\rm ac}} = \frac{\tilde{r}_j}{\Gamma c}\ ,
\end{equation}
which is energy independent.
By solving $t^\prime_{\pi,{\rm ac}} = \tau_\pi
E^\prime_\pi/(m_\pi c^2)$, we find that the neutrino spectrum breaks at
\begin{equation}
E_{\nu,\pi,{\rm ac}} = \frac{b_\pi \Gamma^2}{(1+z)^2} \frac{2 m_\pi c^2 t_v}{\tau_\pi} = 2.7 \times 10^9  \frac{\Gamma_{2.5}^2 t_{v,-2}}{(1+z)^2}\ \mathrm{GeV}\ ,
\label{eq:acpi}
\end{equation}
and it steepens with respect to the parent spectrum by $1/E^{\prime}_{\nu}$.

The radiative cooling time is given by
\begin{equation}
\label{eq:tprimerc}
t^\prime_{\pi,{\rm rc}} = \frac{3 m_\pi^4 c^3}{4 \sigma_T m_e^2 E^\prime_\pi (U^\prime_B+U^\prime_\gamma)} = \frac{3 \pi c^6 m_\pi^4 \Gamma^6 t_v^2}{0.3 \sigma_T m_e^2 E^\prime_\pi \tilde{L}_{\rm iso} (1+z)^2 (1 + \epsilon_B/\epsilon_e)}\ ,
\end{equation}
where $U^\prime_B = B^{\prime 2}/(8 \pi) = 4 \epsilon_B
E^\prime_j/V^\prime$ defined in terms of the fraction of the internal
energy carried by the magnetic field ($\epsilon_B$), $U^\prime_\gamma =
E^\prime_\gamma n^\prime_\gamma = 4 E^\prime_j
\epsilon_e/V^\prime$,\footnote{The numerical factor $4$ in the
definition of $U^\prime_B$ and $U^\prime_\gamma$ comes from the fact
that, assuming that the upstream material is cold while the shock wave
is propagating, the relativistic strong shock transition relations
predict a post-shock energy density $U_2 = 4 \Gamma^2_{21} n_1
\epsilon$, being $\Gamma_{21} \simeq 1$ the relative Lorenz factor for
the internal shock model and  $n_1$ the density before the shock front,
and $\epsilon$ the  energy carried by the
particle~\cite{Meszaros:2006rc}.} $m_e$ the electron mass and $\sigma_T$ the
Thomson cross section. Note that $t^\prime_{\pi,{\rm rc}} \propto
1/E^\prime_\pi$.
When $t^\prime_c \simeq t^\prime_{\pi, {\rm rc}}$ and
the other cooling processes do not play a role at any relevant
energies, a break energy in the neutrino spectrum is expected for
$t^\prime_{\pi,{\rm rc}}= \tau_\pi E^\prime_\pi/(m_\pi c^2)$:
\begin{equation}
E_{\nu,\pi,{\rm rc}} = \frac{b_\pi \Gamma}{(1+z)^2}\ \left[\frac{3 \pi c^8 m_\pi^5 \Gamma^6 t_v^2}{0.3 \tau_\pi \sigma_T m_e^2 \tilde{L}_{\rm iso}  (1  + \epsilon_B/\epsilon_e)}\right]^{1/2}  = 1.8 \times 10^8 \frac{\Gamma_{2.5}^4 t_{v,-2}}{(1+z)^2 [\tilde{L}_{{\rm iso},52}  (1  + \epsilon_B/\epsilon_e)]^{1/2}}\ \mathrm{GeV}\ , 
\label{eq:rcpi}
\end{equation} 
and the radiative cooling is responsible for a steepening of the
correspondent neutrino spectrum by a factor $\propto 1/E^{\prime\
2}_{\nu}$.

If the hadronic cooling occurs at energies lower than the ones for
which the radiative cooling dominates, the transition
between the two cooling mechanisms happens at the neutrino break energy
corresponding to $t^\prime_{\pi,{\rm
hc}}=t^\prime_{\pi,{\rm rc}}$ instead of $t^\prime_{\pi,{\rm rc}}=
\tau_\pi E^\prime_\pi/(m_\pi c^2)$; it is:
\begin{equation}
E_{\nu,\pi,{\rm rc},2} = \frac{b_\pi \Gamma}{(1+z)}\ \frac{3 c^2 m_\pi^4
 \sigma_h}{20 m_p m_e^2 \sigma_T (\epsilon_e + \epsilon_B)}
 = 1.4\times 10^{3} \frac{\Gamma_{2.5}}{(1+z) (\epsilon_e + \epsilon_B)} \mathrm{GeV}\ .
\label{eq:rc2pi}
\end{equation}
Similarly, if the adiabatic cooling occurs before the radiative cooling,  the break energy in the neutrino spectrum due to the
radiative cooling is determined by $t^\prime_{\pi,{\rm ac}}=t^\prime_{\pi,{\rm rc}}$:
\begin{equation}
E_{\nu,\pi,{\rm rc},3} = \frac{b_\pi \Gamma^6}{(1+z)^2} \frac{5 \pi c^6 m_\pi^4 t_v}{(1 + \epsilon_B/\epsilon_e) \tilde{L}_{\rm iso} m_e^2 \sigma_T}  = 1.2 \times 10^7 \frac{\Gamma_{2.5}^6 t_{v,-2}}{(1+z)^2 (1 + \epsilon_B/\epsilon_e) \tilde{L}_{{\rm iso},52}}\ \mathrm{GeV}\ .
\label{eq:rc3pi}
\end{equation}
In both cases the correspondent neutrino spectrum will be subject to a
further $1/E^{\prime}_{\nu}$ steepening with respect to the one affected
by hadronic or adiabatic cooling.

Muons produced by the pion decay (from now on indicated as $\mu_\pi$) will in turn originate a neutrino
energy spectrum, with a first break energy defined similarly to the one
in Eq.~(\ref{eq:firstbreak}), but with the pre-factor $a_{\mu_\pi} =
0.05 = 0.2 \times 3/4 \times 1/3$ (where $3/4$ is the
energy fraction transferred from pions to muons, and
$1/3$ is due to three-body decay of the muon).  In terms of typical GRB parameters, it can be written as
\begin{equation}
E_{\nu,b,{\mu_\pi}} = 1.6 \times 10^6 E_{\gamma, b,{\mathrm{MeV}}}^{-1} \left(\frac{\Gamma_{2.5}}{1+z}\right)^2\ \mathrm{GeV}\ ,
\end{equation}

Other energy breaks in the neutrino  spectrum are determined by  the radiative cooling processes ($t^\prime_{\mu,{\rm rc}} \simeq \tau_\mu E^\prime_\mu/(m_\mu c^2)$):
\begin{equation}
E_{\nu,\mu,{\rm rc}} = \frac{b_{\mu} \Gamma}{(1+z)^2}\ \left[\frac{3 \pi c^8 m_\mu^5 \Gamma^6 t_v^2}{0.3 \tau_\mu \sigma_T m_e^2 \tilde{L}_{\rm iso}  (1  + \epsilon_B/\epsilon_e)}\right]^{1/2} = 1.3 \times 10^7 \frac{\Gamma_{2.5}^4 t_{v,-2}}{(1+z)^2 [\tilde{L}_{{\rm iso},52}  (1  + \epsilon_B/\epsilon_e)]^{1/2}}\ \mathrm{GeV}\ , 
\label{eq:rcmu}
\end{equation} 
with $b_{\mu} = 1/3$ since $E^\prime_\nu = E^\prime_\mu/3$.  Note as the break energy $E_{\nu,\mu,{\rm rc}}$ is usually one order of magnitude smaller than the pion one (Eq.~\ref{eq:rcpi}), due to difference in the masses and lifetimes of the two particles. Similarly to pions, muons are also subject to adiabatic cooling. In this case, the correspondent break energies in the neutrino spectrum are
\begin{eqnarray}
\label{eq:acmu}
E_{\nu,\mu,{\rm ac}} &=& \frac{b_\mu \Gamma^2}{(1+z)^2} \frac{2 m_\mu c^2 t_v}{\tau_\mu}  = 3.2 \times 10^7 \frac{\Gamma_{2.5}^2 t_{v,-2}}{(1+z)^2}\ \mathrm{GeV}\ ,\\ 
E_{\nu,\mu,{\rm rc},3} &=& \frac{b_\mu \Gamma^6}{(1+z)^2} \frac{5 \pi c^6 m_\mu^4 t_v}{(1 + \epsilon_B/\epsilon_e) \tilde{L}_{\rm iso} m_e^2 \sigma_T}  = 5.4 \times 10^6 \frac{\Gamma_{2.5}^6 t_{v,-2}}{(1+z)^2 (1 + \epsilon_B/\epsilon_e) \tilde{L}_{{\rm iso},52}}\ \mathrm{GeV}\ .
\label{eq:rc3mu}
\end{eqnarray}

\begin{figure}[!t]
\begin{center}  
\hspace{-2mm}\includegraphics[width=0.6\columnwidth]{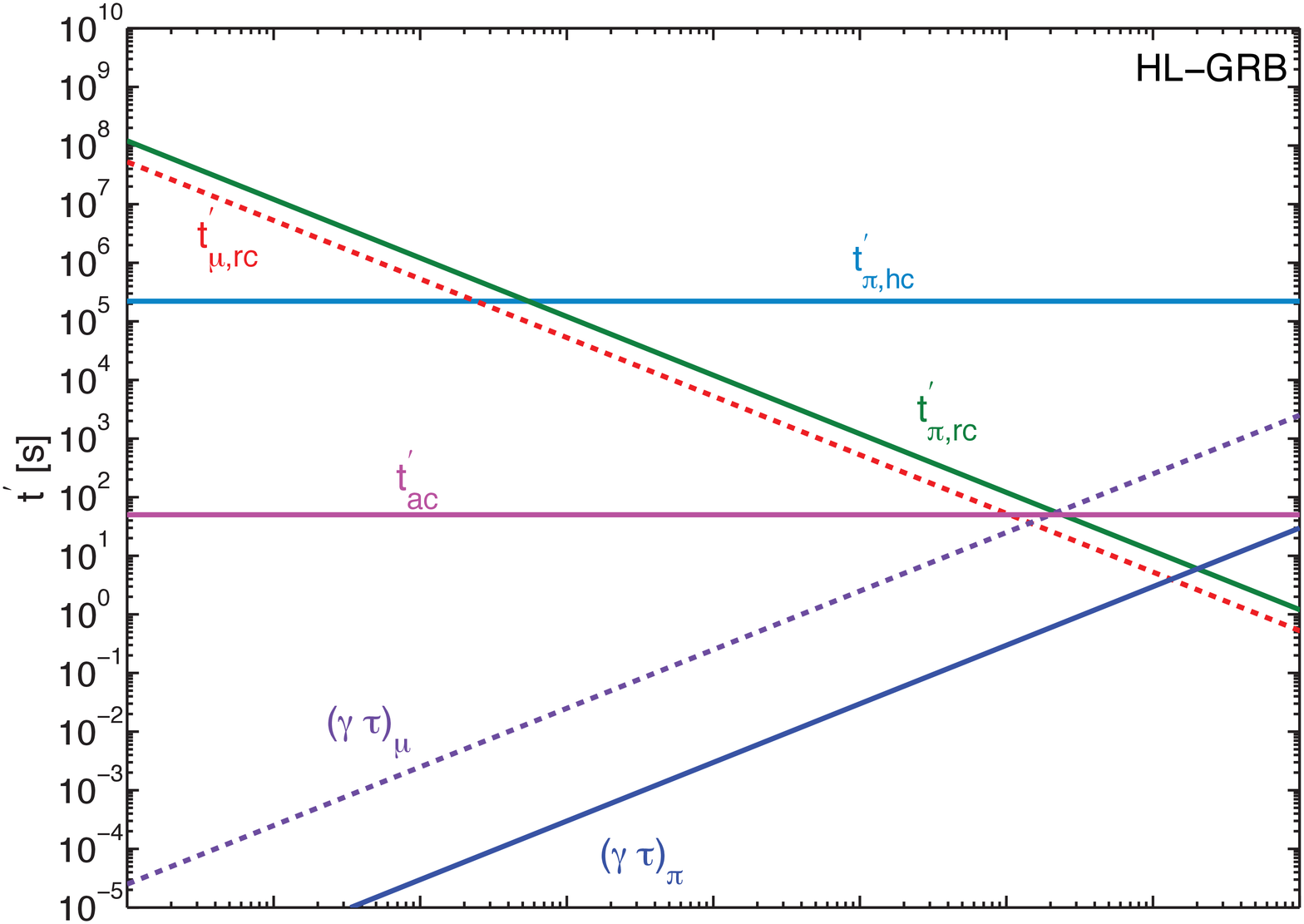} \\
\includegraphics[width=0.61\columnwidth]{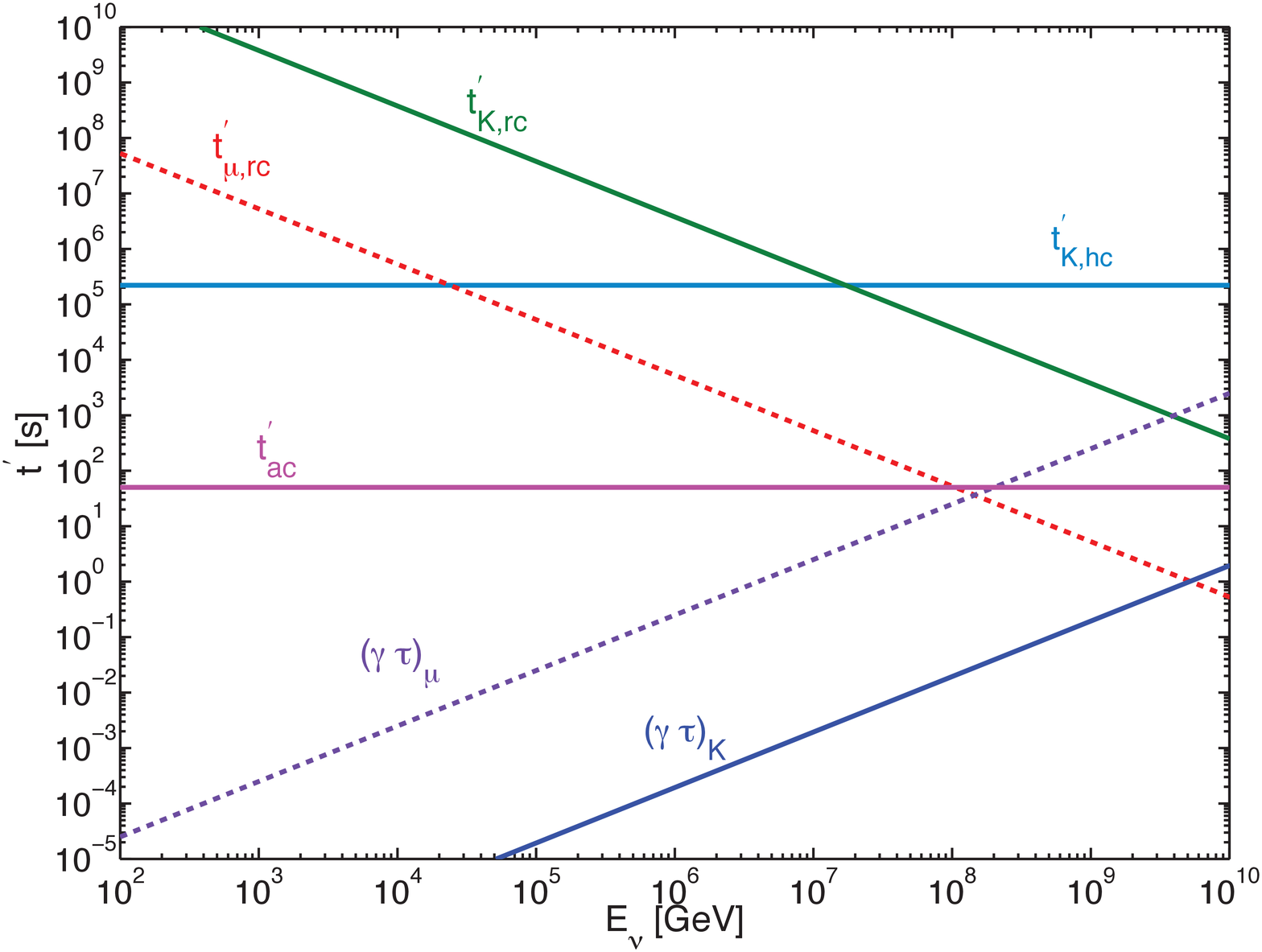}  
\end{center}  
\caption{Top panel: Muon and pion lifetimes and cooling times in the
jet comoving frame as a function of the neutrino energy $E_\nu$ for a typical HL-GRB with $\tilde{L}_{\rm iso} = 10^{52}~\mathrm{erg}/{\rm s}$ and $z = 1$. Bottom panel: Muon and kaon lifetimes and cooling times  for the same HL-GRB. For the assumed HL-GRB parameters, the radiative cooling is always important, while the hadronic cooling is negligible and the adiabatic cooling is relevant for muons. \label{fig:timespiHL}}  
\end{figure}  
\begin{figure}
\begin{center}  
\hspace{-2mm}\includegraphics[width=0.6\columnwidth]{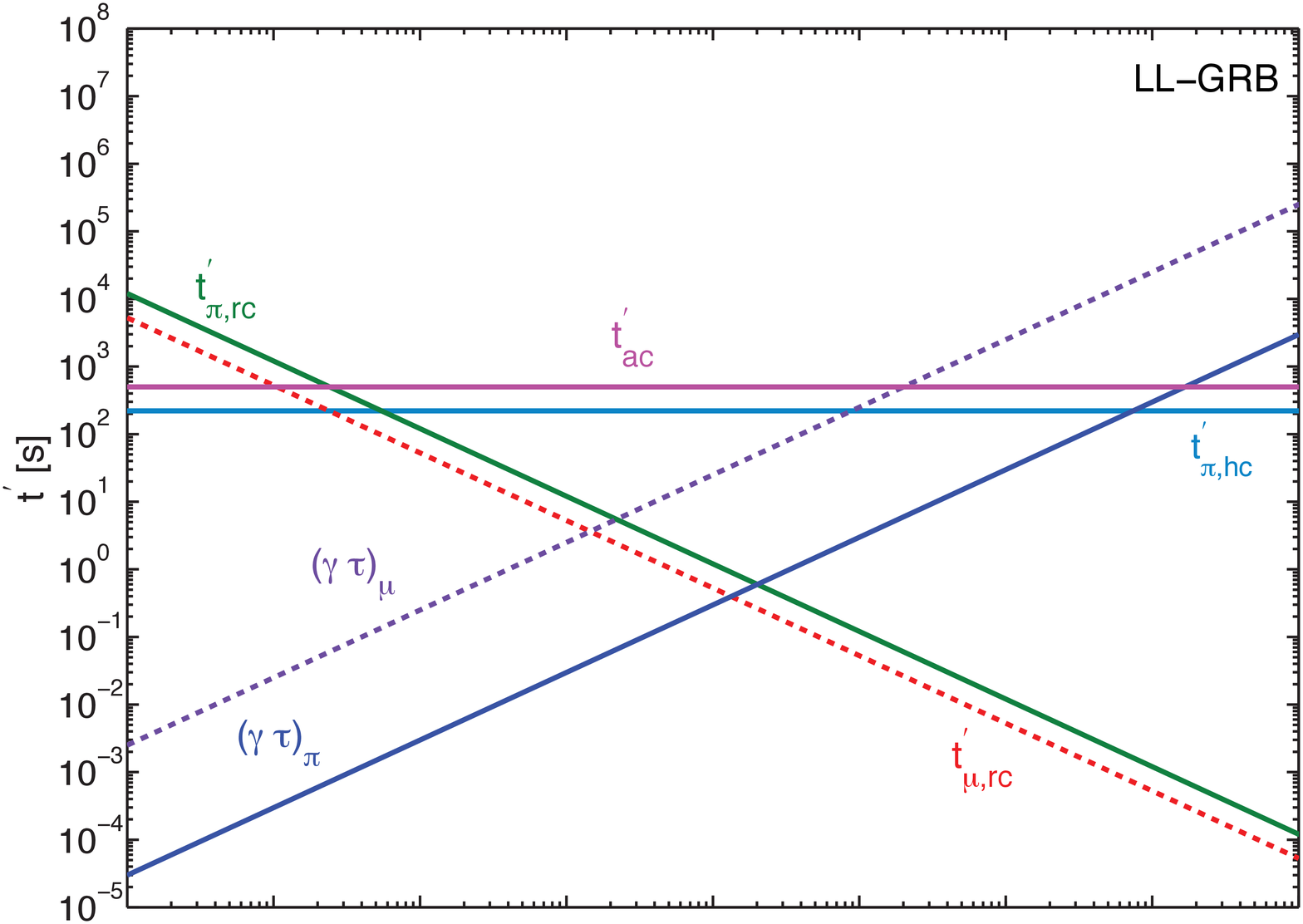} \\
\includegraphics[width=0.61\columnwidth]{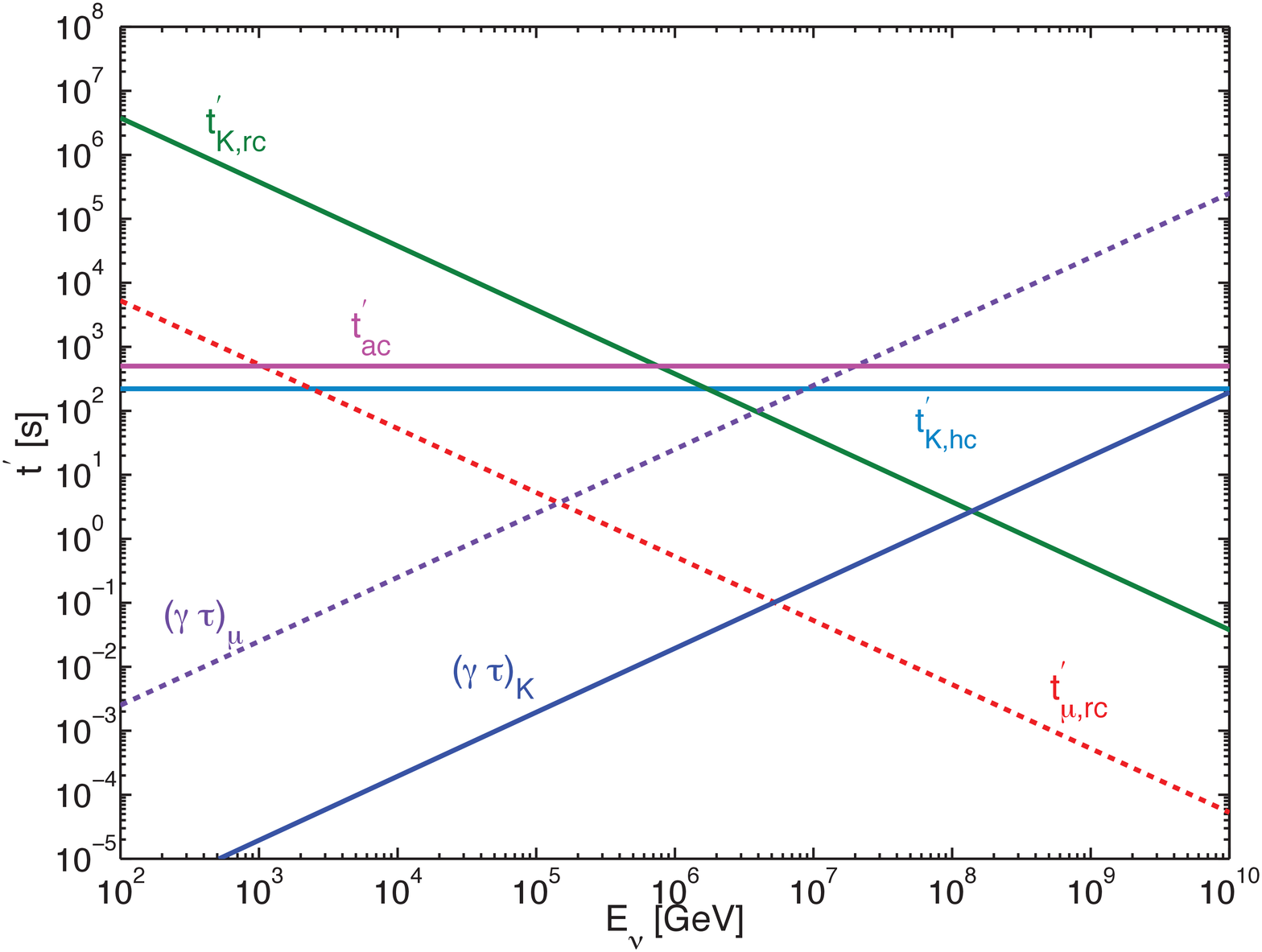}  
\end{center}  
\caption{The same as Fig.~\ref{fig:timespiHL}, but for  a typical LL-GRB with $\tilde{L}_{\rm iso} = 10^{48}~\mathrm{erg}/{\rm s}$ and $z = 1$. For the assumed LL-GRB parameters,  the adiabatic and the hadronic cooling are negligible for kaons. \label{fig:timespiLL}}  
\end{figure}  
\begin{figure}
\begin{center}  
\includegraphics[width=0.6\columnwidth]{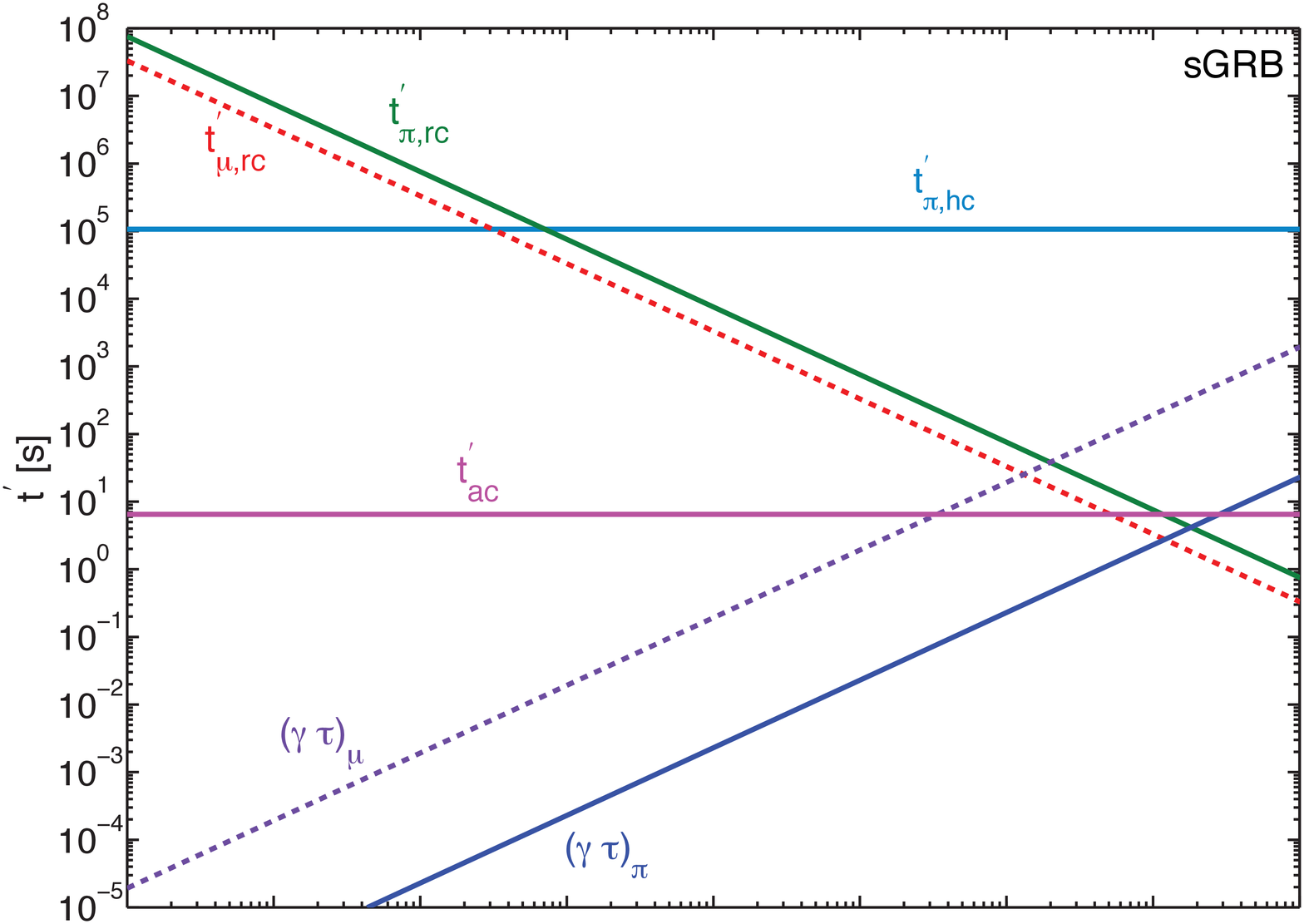} \\
\hspace{2mm}\includegraphics[width=0.61\columnwidth]{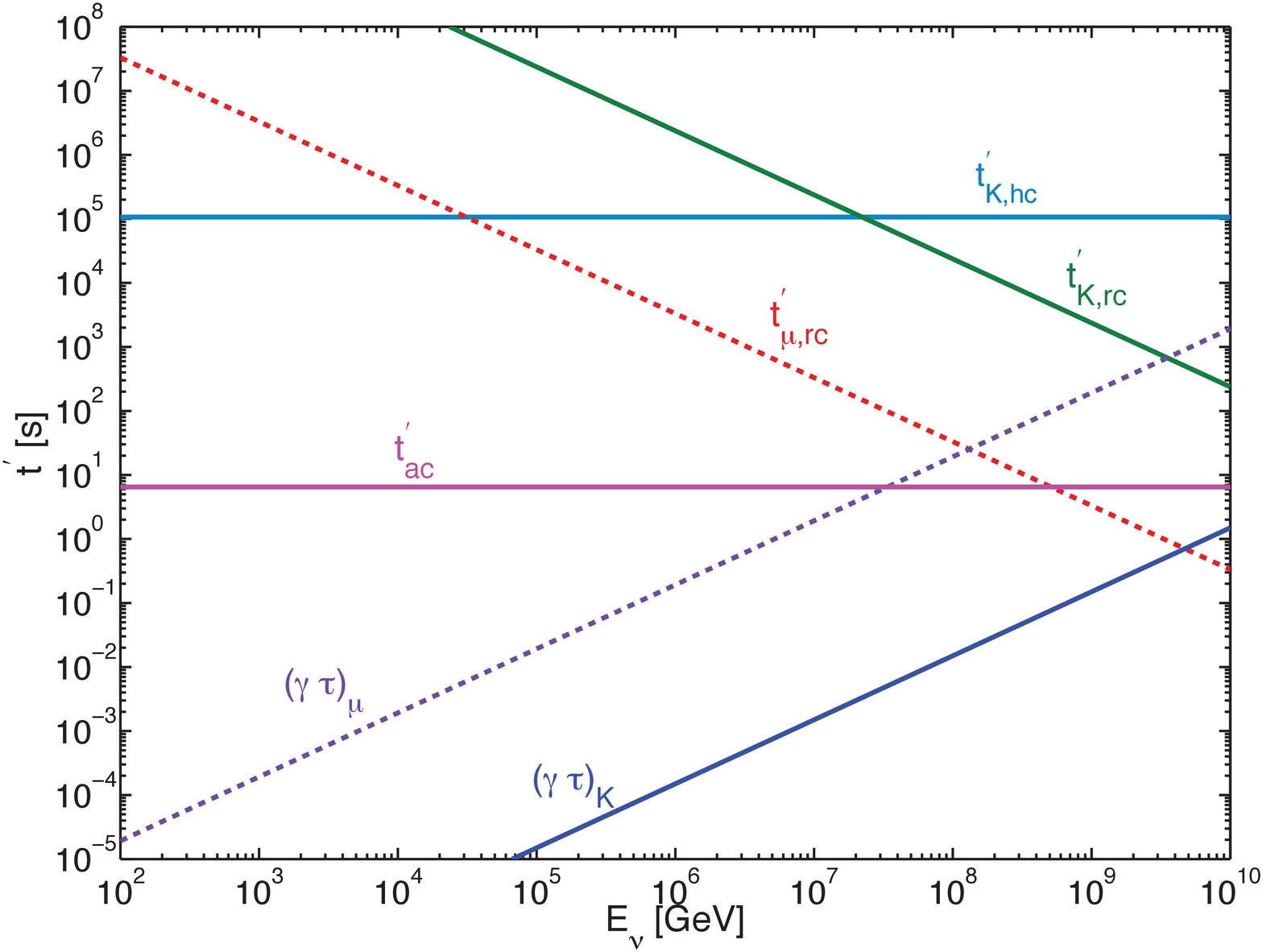}  
\end{center}  
\caption{The same as Fig.~\ref{fig:timespiHL}, but for  a typical sGRB with $\tilde{L}_{\rm iso} = 10^{51}~\mathrm{erg}/{\rm s}$ and $z = 1$. For the assumed sGRB parameters, the hadronic cooling is negligible.\label{fig:timespis}}  
\end{figure}  

Assuming that $E^\prime_\pi = 4 E_\nu (1+z)/\Gamma$ and $E^\prime_\mu =
3 E_\nu (1+z)/\Gamma$, the relations among the comoving muon and pion
lifetimes and the cooling times as from Eqs.~(\ref{eq:tprimehc}),
(\ref{eq:tprimeac}), (\ref{eq:tprimerc}) and the correspondent ones for
muons are shown in the top panels of Figs.~\ref{fig:timespiHL},
\ref{fig:timespiLL} and \ref{fig:timespis} for typical HL-, LL-GRB and
sGRBs  at $z = 1$. We took $\tilde{L}_{\rm iso} =
10^{52}~\mathrm{erg}/{\rm s}$  for the HL-GRBs, $\tilde{L}_{\rm iso} =
10^{48}~\mathrm{erg}/{\rm s}$ for the LL-GRB, and $\tilde{L}_{\rm iso} =
10^{51}~\mathrm{erg}/{\rm s}$ for the sGRBs. For each family, we supposed
$\Gamma$ and $t_v$ as in Table~\ref{table:LF} and defined   $\epsilon_e =
\epsilon_B = 10^{-2}$ for HL-GRBs and sGRBs, while we adopted $\epsilon_e =
\epsilon_B = 10^{-3}$ for LL-GRBs (see Sec.~\ref{sec:spectra}). 

For the GRB parameters assumed in Figs.~\ref{fig:timespiHL},
\ref{fig:timespiLL} and \ref{fig:timespis}, the hadronic cooling is
negligible for HL-GRBs and sGRBs. On the other hand,  the
radiative cooling and the adiabatic one are always relevant for all three families for both
pions and muons. As we will discuss in the following, note as such
hierarchy among the cooling processes is a function of $\tilde{L}_{\rm
iso}$ and $z$ for other fixed GRB parameters.  Therefore it will change
within the luminosity and redshift range that we will consider for the
computation of the diffuse neutrino emission.

\subsection{Neutrino production from kaon decay}

Yet another contribution to the total neutrino spectrum from GRBs
originates from kaon decays. The resultant neutrino spectrum will have a
first break energy coming from the proton threshold energy for kaon
production, similarly to Eq.~(\ref{eq:firstbreak}):
\begin{eqnarray}
E_{\nu,b,i} &=& c_i \left(\frac{\Gamma}{1+z}\right)^2 \frac{(m_K c^2 + m_\Lambda c^2)^2-(m_p c^2)^2}{2 E_{\gamma,b}} \ ,
\end{eqnarray}
where, for neutrinos directly produced from kaon decay
without going through muons, $c_{K} = 0.1$ (since $20\%$ is the
fraction of the proton energy that goes into $K$ and $1/2$ is the
fraction of $K$ energy carried by neutrinos). In the case of muons
originating from kaon decay, from now on indicated as $\mu_K$ to distinguish them from the ones from muon decay ($\mu_\pi$), $c_{\mu_K} = 0.033 = 0.2 \times 1/2 \times
1/3$.  In terms of typical parameters, the first break energies for $K$ and $\mu_K$ become:
\begin{eqnarray}
E_{\nu,b,K} &=&  8.5 \times 10^6 E_{\gamma, b,{\mathrm{MeV}}}^{-1} \left(\frac{\Gamma_{2.5}}{1+z}\right)^2\ \mathrm{GeV}\ ,\\
E_{\nu,b,\mu_K} &=& 2.6 \times 10^6 E_{\gamma, b,{\mathrm{MeV}}}^{-1} \left(\frac{\Gamma_{2.5}}{1+z}\right)^2\ \mathrm{GeV}\ .
\end{eqnarray}

The resultant neutrino spectrum from kaon decay is affected, similarly
to what was discussed for pions, from hadronic, adiabatic, and radiative
cooling processes. 
Therefore, we should expect breaks in the neutrino spectrum similar to the ones described in Eqs.~(\ref{eq:hcpi}), (\ref{eq:acpi}), (\ref{eq:rcpi}), (\ref{eq:rc2pi}), and (\ref{eq:rc3pi}) according to the cases, but with the substitution
$m_\pi(\tau_\pi) \rightarrow m_K(\tau_K)$. For example, $E_{\nu,K,{\rm
rc}}$ will be $d_K \Gamma/(1+z)^2\ [4 \pi c^8 m_\pi^5 \Gamma^6 t_v^2/(0.3 \tau_\pi \sigma_T m_e^2 \tilde{L}_{\rm iso}  (1  + \epsilon_B/\epsilon_e))]^{1/2}$, 
with $d_K = 1/2$ since $E^\prime_\nu = E^\prime_K/2$. 
Similarly, the equations in terms  of typical parameters can be extrapolated from the ones from pions rescaling  
$m_\pi(\tau_\pi)$ to $m_K(\tau_K)$ and taking into account the different multiplicity factors.

Muons are also produced from kaon decay. The resultant neutrino energy
spectrum will be determined as  the one described for muons from pion
decay (see Eqs.~\ref{eq:rcmu}, \ref{eq:acmu} and \ref{eq:rc3mu}).

The bottom panels of Figs.~\ref{fig:timespiHL}, \ref{fig:timespiLL} and
\ref{fig:timespis} show the relations among the muon and kaon lifetimes
and the cooling times, similarly to the top panels for pions and muons
from pion decays, assuming that $E^\prime_K = 2 E_\nu (1+z)/\Gamma$ and
$E^\prime_\mu = 3 E_\nu (1+z)/\Gamma$.  The hadronic cooling is
negligible, while the
 adiabatic cooling is relevant for muons for the
model parameters plotted in these figures. Note that  the break energies
in the neutrino spectrum due to the kaon radiative cooling occur  at
higher energies than the ones due to pion radiative cooling given the
differences in the rest-mass and lifetimes of the two parent
particles~\cite{Ando:2005xi, Asano:2006zzb}.  We neglect here the contribution to the neutrino flux 
coming from $K^0$, see Ref.~\cite{Petropoulou:2014lja} for a dedicated discussion.

\subsection{Neutrino energy spectra}\label{sec:spectra}

As discussed in the previous section, the neutrino energy spectrum for
one flavor ($\nu + \bar{\nu}$) resulting from $\pi$, $K$, and $\mu$
decays will be a broken power law derived from the parent proton
spectrum (that we assume is proportional to $E_p^{-2}$) with further breaks
defined according to the hierarchy of the cooling processes.  For
example, when the pion hadronic cooling is negligible, the adiabatic
cooling is relevant for muons only, and $E^{\prime}_{\nu,b,\mu} <
E^{\prime}_{\nu,\mu,{\rm ac}}< E^{\prime}_{\nu,\mu,{\rm rc,3}} <
E^{\prime}_{\nu,\pi,{\rm rc}}$, the resultant neutrino energy spectrum
produced from muons from pion decay will be:
\begin{eqnarray}
\label{eq:gammaspectrum}
\left(\frac{dN_{\nu}}{dE^\prime_\nu}\right)_{\rm inj, \mu_\pi} \propto  \left\{ \begin{array}{ll}
  \left(\frac{E^\prime_\nu}{E^{\prime}_{\nu,b,\mu}}\right)^{\beta_\gamma - 3}&   \mathrm{for}\ \ E^\prime_\nu < E^{\prime}_{\nu,b,\mu} \\ 
   \left(\frac{E^{\prime}_\nu}{E^{\prime}_{\nu,b,\mu}}\right)^{\alpha_\gamma - 3}&  \mathrm{for}\ \  E^{\prime}_{\nu,b,\mu} \le E^\prime_\nu < E^\prime_{\nu,\mu,{\rm ac}}\\
  \left(\frac{E^\prime_{\nu,\mu,{\rm ac}}}{E^{\prime}_{\nu,b,\mu}}\right)^{\alpha_\gamma-3}\left(\frac{E^\prime_{\nu}}{E^\prime_{\nu,\mu,{\rm ac}}}\right)^{\alpha_\gamma - 4}& \mathrm{for}\ \  E^{\prime}_{\nu,\mu,{\rm ac}} \le E^\prime_\nu < E^\prime_{\nu,\mu,{\rm rc},3}\\
  \left(\frac{E^\prime_{\nu,\mu,{\rm ac}}}{E^{\prime}_{\nu,b,\mu}}\right)^{\alpha_\gamma-3}\left(\frac{E^\prime_{\nu,\mu,{\rm rc,3}}}{E^\prime_{\nu,\mu,{\rm ac}}}\right)^{\alpha_\gamma-4}\left(\frac{E^\prime_\nu}{E^\prime_{\nu,\mu,{\rm rc,3}}}\right)^{\alpha_\gamma - 5}&   \mathrm{for}\  \  E^\prime_{\nu,\mu,{\rm rc,3}} \le E^\prime_\nu <  E^\prime_{\nu,\pi,{\rm rc}}\\
 \left(\frac{E^\prime_{\nu,\mu,{\rm ac}}}{E^{\prime}_{\nu,b,\mu}}\right)^{\alpha_\gamma-3}\left(\frac{E^\prime_{\nu,\mu,{\rm rc,3}}}{E^\prime_{\nu,\mu,{\rm ac}}}\right)^{\alpha_\gamma-4}\left(\frac{E^\prime_{\nu,\pi,{\rm rc}}}{E^\prime_{\nu,\mu,{\rm rc,3}}}\right)^{\alpha_\gamma-5}\left(\frac{E^\prime_{\nu}}{E^\prime_{\nu,\pi,{\rm rc}}}\right)^{\alpha_\gamma - 7}&   \mathrm{for}\  \  E^\prime_\nu \ge  E^\prime_{\nu,\pi,{\rm rc}}\\ 
\end{array}\right .\nonumber\\
\end{eqnarray}

For each decay channel $i$, the neutrino energy spectrum is normalized in terms of the total photon fluence by generalizing the expression proposed in Refs.~\cite{Li:2011ah,Hummer:2011ms}: 
\begin{equation}
\int_0^\infty  dE_\nu E_\nu \left(\frac{dN_\nu}{dE_\nu}\right)_{{\rm{inj}},i} = N_i \frac{h_{p,i}}{h_{\gamma p}} [1 - (1 - \langle \chi_{p}\rangle)^{\tau_{p\gamma}}]  \int_{0}^{\infty} dE_\gamma E_\gamma \left(\frac{dN_\gamma}{dE_\gamma}\right)_{\rm inj} \ .
\label{eq:normpi}
\end{equation}
In our numerical computations within the \emph{canonical} model, we assume  the
gamma-ray--proton luminosity ratio $h_{\gamma p} = L_{\rm iso}/L_p =
10^{-2}$ for HL-GRBs and sGRBs as suggested by  joint analysis of
high-energy neutrino and ultrahigh-energy cosmic ray (UHECR) data~\cite{Asano:2014nba,Baerwald:2014zga}  and 
under the assumption that sGRBs behave similarly to HL-GRBs; while we adopt $h_{\gamma p} = 10^{-3}$ for the LL-GRB family assuming that the
$\gamma$ production is suppressed in these GRBs with respect to the other two GRB families for the same $L_p$\footnote{Note as, assuming
$L_j = L_p+L_e+L_B$ and $\epsilon_e \simeq \epsilon_B$, one has that $h_{\gamma p} \simeq \epsilon_e/(1-\epsilon_e-\epsilon_B) \sim \epsilon_e$. This justifies 
the choice of the numerical values of $\epsilon_e$ and $\epsilon_B$ introduced in Sec.~\ref{sec:cooling} and adopted trough  the whole paper.}.
 We will discuss in detail the dependence of the diffuse background from this parameter in Sec.~\ref{sec:parameters}. 
For the case of neutrinos produced from the pion decay, $N_\pi = 0.97/8$
(the coefficient $1/8 = 1/4 \times 1/2$  since each neutrino takes about
$1/4$ of the pion energy and $1/2$ of the produced pions are charged
pions, while $97\%$ is the probability that pions are produced from one
$p\gamma$ interaction~\cite{Kelner:2008ke,Mucke:1999yb}) and
 $h_{p,\pi} \simeq \ln(E_{\nu,\pi,{\rm second}}/E_{\nu,\pi,{\rm first}})/\ln(E_{p, {\rm max}}/E_{p, {\rm min}})$ with $E_{\nu,\pi,{\rm first}}$ the minimum neutrino break energy and $E_{\nu,\pi,{\rm second}}$ the second neutrino break energy, both  defined according to the hierarchy  of the break energies determined by the cooling processes and $E_{\nu,b,\mu}$. Note that, for each GRB family and fixed parameters, such energy hierarchy varies as a function of  $\tilde{L}_{\rm iso}$ and $z$. The minimum proton energy is $E_{p, {\rm min}} = \Gamma m_p c^2/(1+z)$  while the maximum proton energy is $E_{p, {\rm max}}$ and it is defined 
as the energy when the acceleration time $t^\prime_{p,{\rm acc}} =
E^\prime_p/(B^\prime e c)$ equals to $t^\prime_{p,c} = {\rm min}(t^\prime_{p,{\rm sync}}, t^\prime_{p,{\rm dyn}})$, being  $t^\prime_{p,{\rm sync}} = (3 m_p^4 c^3 8 \pi)/(4 \sigma_T m_e^2 E^\prime_p B^{\prime 2})$ the synchrotron cooling time and $t^\prime_{p,{\rm dyn}} = r^\prime_j/c$ the dynamical time scale.
For $\mu$ produced from $\pi$ decay, $N_{\mu_\pi} = 0.97/3 \times 3/8 = N_{\pi}$ and $h_{p,\mu_\pi} \simeq \ln(E_{\nu,\mu,{\rm second}}/E_{\nu,\mu,{\rm first}})/ \ln(E_{p, {\rm max}}/E_{p, {\rm min}})$ defined analogously to $h_{p,\mu}$. 

In the case of neutrino production from kaon decay, $N_K = 0.03 \times
0.63/2$ where the numerical coefficient  $1/2$  comes from the fact that
each neutrino takes about $1/2$ of the kaon's energy, $0.63$ is the
probability that $K$ decays in neutrinos, and $3\%$ the probability that
kaons are produced from one $p\gamma$
interaction~\cite{Kelner:2008ke,Mucke:1999yb}. The term $h_{p,K} \simeq
\ln(E_{\nu,K,{\rm second}}/E_{\nu,K,{\rm first}})/\ln(E_{p, {\rm
max}}/E_{p, {\rm min}})$ is defined similarly as $h_{p,\pi}$.
For muons generated from kaon decay, we have
$N_{\mu_K} = 0.03 \times 0.63/6$.

The second term in Eq.~(\ref{eq:normpi}), $f = [1 - (1 - \langle
\chi_{p}\rangle)^{\tau_{p\gamma}}]$, represents the fraction of the
proton energy that goes to pion production with $\langle
\chi_{p}\rangle \simeq 0.2$ the average fraction of energy transferred
from protons to pions or kaons per $p\gamma$
interaction~\cite{Hummer:2011ms,Zhang:2012qy}. The $p\gamma$ optical
depth, $ \tau_{p\gamma}$, is defined
following~\cite{Waxman:1997ti,Waxman:2003vh,Zhang:2012qy}:
 \begin{equation}
 \tau_{p\gamma} = \frac{\tilde{r}_j}{\Gamma \ell_{p\gamma}} = 0.6 \frac{\tilde{L}_{\rm iso}}{10^{52}\ {\rm erg}/{\rm s}} \left(\frac{\Gamma}{10^{2.5}}\right)^{-2} \left(\frac{\tilde{r}_j}{10^{14} {\rm cm}}\right)^{-1}\ , 
  \label{eq:taupgamma}
  \end{equation}
 with $\ell_{p\gamma}$ the mean free path for $p\gamma$ interactions.
  Note that with respect to the numerical fireball model revised  in \cite{Hummer:2011ms}, we do include  their  correction term $c_s$
 (originally introduced in \cite{Li:2011ah}), but  do not implement 
  the factor $c_{f_\pi}$ due to the pion efficiency (see \cite{Hummer:2011ms} for details). The latter is responsible
  for a further reduction of the expected neutrino flux with respect to the one obtained  from our Eq.~(\ref{eq:normpi}), as shown in the left panel of Fig.~1 of 
  \cite{Hummer:2011ms}. We also neglect multi-pion processes that should be instead responsible for an enhancement of the neutrino flux (see
   right  panel of Fig.~1 of Ref.~\cite{Hummer:2011ms}, comparison between their RFC and NFC models). Therefore our analytical model is still an approximation with respect to the full numerical treatment, but it gives us results in good agreement with the revised fireball model of ~\cite{Hummer:2011ms} for the same choice of the initial parameters.

In order  to consistently normalize the neutrino spectrum as in 
Eq.~(\ref{eq:normpi}), we need to select the  GRB parameters
in such a way to have optically thin media~\cite{Waxman:1997ti}. Therefore, 
we consider a  test photon with energy $E_{\gamma,t} = 100$~MeV  and
define the $\gamma\gamma$ optical depth 
\begin{equation}
\label{eq:taugammagamma}
\tau_{\gamma\gamma} = \frac{\tilde{r}_j}{\Gamma \ell_{\gamma\gamma}} = \frac{(1+z)^2 \sigma_T 0.3 \tilde{L}_{\rm iso} E_{\gamma,t}}{128 \pi \Gamma^6 c^2 t_v (m_e c^2)^2}\ ,
\end{equation}
with the mean free path for pair production
$\ell_{\gamma\gamma}^{-1} \simeq (\sigma_T U^{\prime}_{\gamma}
E^{\prime}_{\gamma,t})/[16 (m_e c^2)^2]$ and $\tilde{L}_{\rm ave} =
0.3\tilde{L}_{\rm iso}= 4 \pi \tilde{r}_j^2 \Gamma^2 c
U^{\prime}_{\gamma}$~\cite{Waxman:2003vh,Waxman:1997ti}. 
 In our {\emph{canonical}} GRB model, we fix $t_v$ to the average value preferred by observations~\cite{MacLachlan:2012cd,Dermer:2014vaa,Liu:2011cua} and  define $\Gamma$   as the 
minimum Lorenz factor required to avoid high pair-production optical depth that would inhibit
 the gamma-ray emission (i.e., $\tau_{\gamma\gamma} \le 1$) for the whole $[\tilde{L}_{\rm iso}, z]$ parameter space of the HL-GRB and sGRB families considered in Sec.~\ref{sec:nubackastro}. The Lorenz factors  $\Gamma$ defined in this way   are reported in Table~\ref{table:LF}. 
Note that $E_{\gamma,t}$ 
 belongs to the upper tail of the observed photon energy spectrum and therefore it allows us to define an upper limit for the optical depth for pair production for the assigned GRB parameters.
The $\tau_{\gamma\gamma} \le 1$ condition does not apply to  LL-GRBs as the average photon energy  is lower than the pair production threshold (see Sec.~\ref{sec:observations}). 
Variations of the expected diffuse neutrino background as a function of $t_v$ and $\Gamma$ will be discussed in Sec.~\ref{sec:parameters}.

 The observed neutrino spectrum from a single source at redshift $z$ is defined as
\begin{equation}
F_{\nu}(E_\nu) = \frac{(1+z)^3}{4 \pi \Gamma d_L^2(z)} \left(\frac{dN_{\nu}}{dE^\prime_\nu}\right)\ ,
\label{eq:siglesource}
\end{equation} 
with $E^\prime = E (1+z)/\Gamma$ and $d_L(z)$ the luminosity distance~\cite{Hogg:1999ad} computed assuming a flat $\Lambda$CDM cosmology with $\Omega_m = 0.32$, $\Omega_\Lambda = 0.68$ and $H_0 = 67$~km s$^{-1}$ Mpc$^{-1}$ for the Hubble constant~\cite{Ade:2013zuv}.
The total $\nu_e$ and $\nu_\mu$ neutrino spectra  from pion decay at the
source and without flavor oscillation are:
$(dN_{\nu_{e}}/dE^\prime_\nu)_{{\rm inj},\pi} =
(dN_{\nu}/dE^\prime_\nu)_{\mu_\pi}$ and
$(dN_{\nu_\mu}/dE^\prime_\nu)_{{\rm inj},\pi} =
(dN_\nu/dE^\prime_\nu)_{\mu_\pi} + (dN_\nu/dE^\prime_\nu)_{\pi}$ and
similarly for kaons. Note that no $\tau$ neutrinos are produced.

\begin{figure}[h!]
\begin{center}  
\includegraphics[width=0.6\columnwidth]{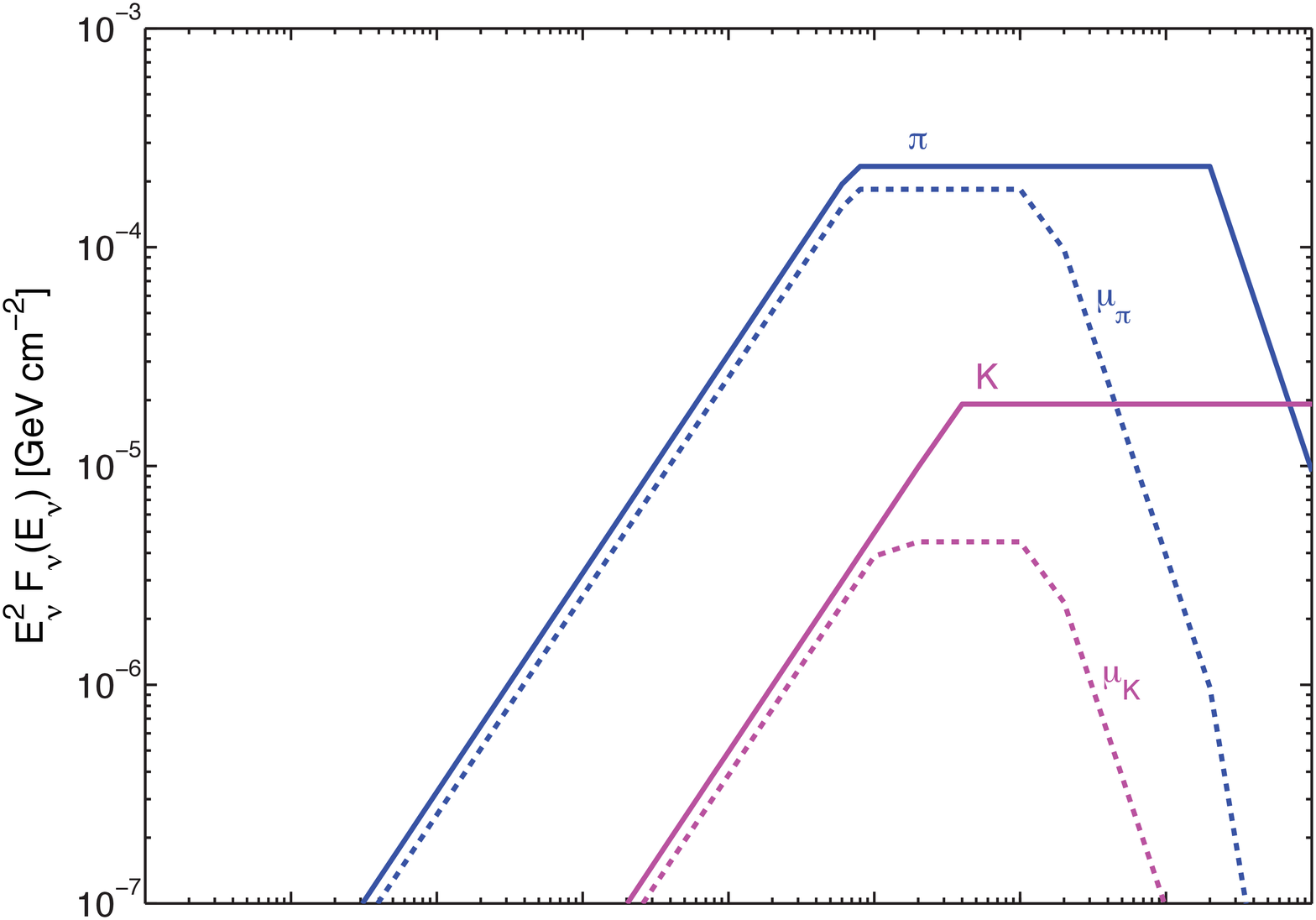} \\ 
\hspace{2mm}\includegraphics[width=0.61\columnwidth]{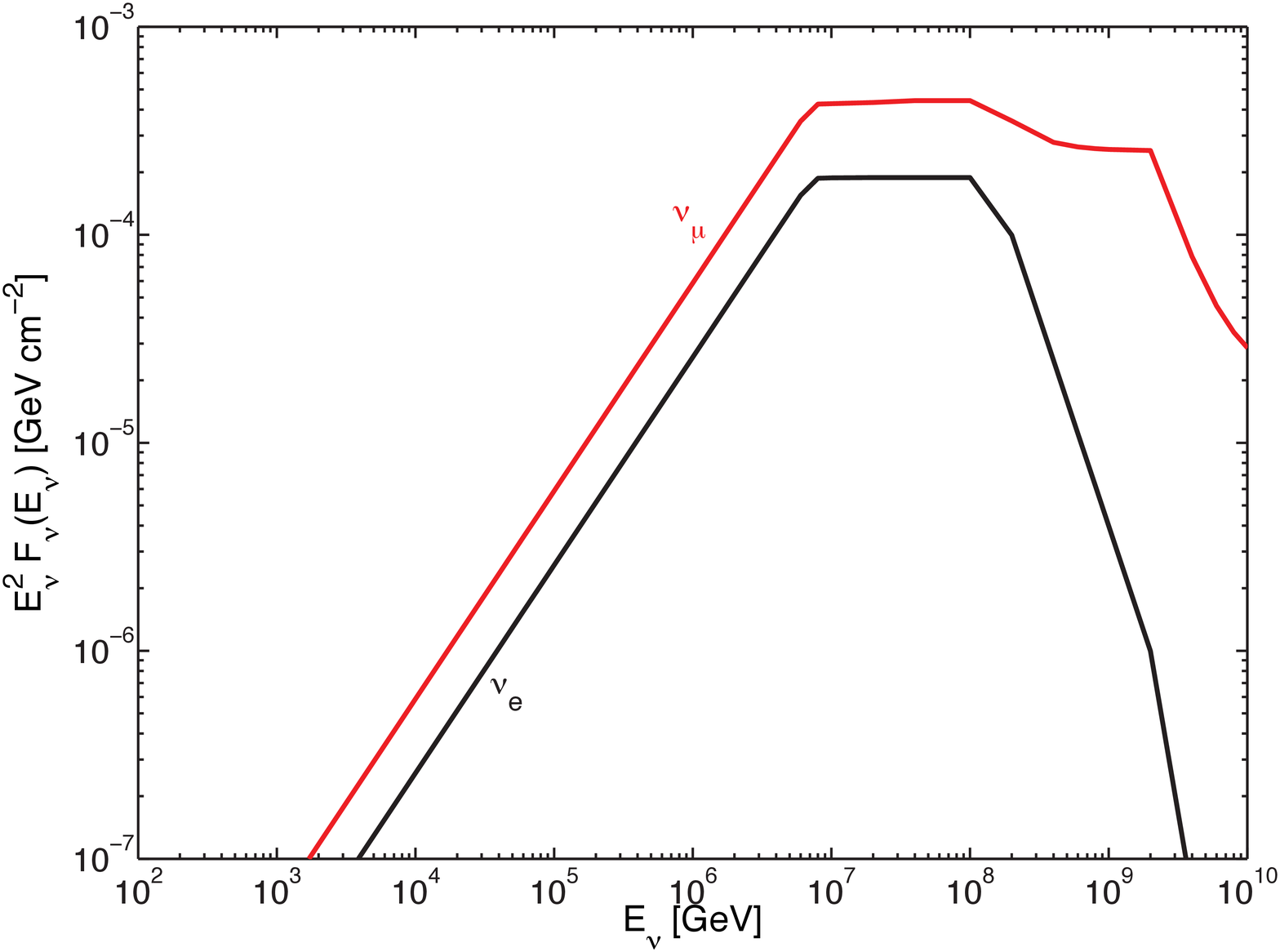}
\end{center}  
\caption{Predicted $E_\nu^2 F_\nu(E_\nu)$ for  a typical HL-GRB
 ($\tilde{L}_{{\rm iso}} = 10^{52}\ {\rm erg}~{\rm s}^{-1}$, $z = 1$)
 without flavor oscillations. Top: Neutrino fluence from $\pi$ (blue
 line) and $K$ (magenta line) decays as well as from $\mu$ from pion
 decay ($\mu_\pi$, dashed blue line) and $\mu$ from kaon decay ($\mu_K$,
 dashed magenta line). Bottom: $E_\nu^2 F_{\nu_e}(E_\nu)$ (black line)
 and $E_\nu^2 F_{\nu_\mu}(dE_\nu)$ (red line)  for  a typical HL-GRB
 and without flavor oscillations.
\label{fig:numunoosci}}  
\end{figure}

Neglecting flavor oscillations, Fig.~\ref{fig:numunoosci} (top panel)
  shows the  neutrino energy spectra coming from pion, kaon and muon
  decays as a function of the energy for a typical HL-GRB  and
  normalized as in Eq.~(\ref{eq:siglesource}) for a source at $z = 1$. 
  The neutrino spectrum coming from muon decay from pions ($\mu_\pi$, dashed blue line) exhibits four breaks, 
 the first one corresponds to the first energy break ($E_{\nu,b,\mu}$),  the second is due to the radiative cooling ($E_{\nu,\mu,{\rm rc}}$), the third to the adiabatic cooling of muons ($E_{\nu,\mu,{\rm ac}}$) and the fourth occurs at the same energy of the second break of the pion spectrum (blue line) and it is indeed due to the radiative cooling of the parent  pion ($E_{\nu,\pi,{\rm rc}}$, see Fig.~\ref{fig:timespiHL} for comparison).  
 For kaons (magenta lines) only the first break energy is relevant, while the neutrino spectrum coming from muon decay from kaons ($\mu_K$, magenta dashed line) is similar to the $\mu_\pi$ one.
The bottom panel of Fig.~\ref{fig:numunoosci}  shows the resultant  $\nu_e$ (in black) and $\nu_{\mu}$ (in red) neutrino energy spectra from pion and kaon decays and as a function of the energy without flavor oscillations for a HL-GRB at $z=1$. 
Note as the cutoff in the neutrino energy spectrum corresponding to the maximum proton energy appears for energies slightly larger than the ones  studied here.

\subsection{Neutrino  flavor oscillations}

While neutrinos travel to reach the Earth they are subject to flavor oscillations and therefore the observed flux will be (see e.g., Appendix B of~\cite{Anchordoqui:2013dnh}):
\begin{eqnarray}
\label{eq:nuoscill}
\left(\frac{dN_{\nu_\mu}}{dE^\prime_\nu}\right)_{\rm osc} &=& \left[\frac{1}{4} \sin^2 (2 \theta_\odot)\right] \left(\frac{dN_{\nu_e}}{dE^\prime_\nu}\right)_{\rm inj} + \frac{1}{8}\left[4 - \sin^2 (2 \theta_\odot)\right]  \left(\frac{dN_{\nu_\mu}}{dE^\prime_\nu}\right)_{\rm inj}\ .
\end{eqnarray}
with $\sin^2 (2 \theta_\odot) \simeq 8/9$. Note that while $\nu_\tau$ are
not produced at the source, the three neutrino flavors are equally
abundant after  flavor oscillations.

\begin{figure}[h!]
\begin{center}  
\includegraphics[width=0.7\columnwidth]{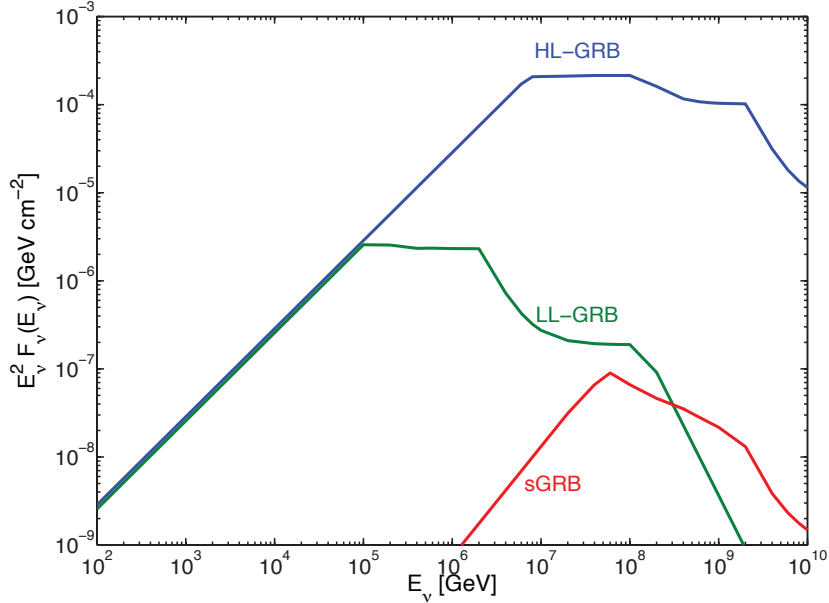} 
\end{center}  
\caption{Predicted $E^2_\nu F^{\nu_\mu}(E^\nu)$ for a  typical HL-GRB  ($\tilde{L}_{{\rm iso}} = 10^{52}\ {\rm erg}~{\rm s}^{-1}$), LL-GRB  ($\tilde{L}_{{\rm iso}} = 10^{48}\ {\rm erg}~{\rm s}^{-1}$), and sGRB ($\tilde{L}_{{\rm iso}} = 10^{51}\ {\rm erg}~{\rm s}^{-1}$) at $z=1$ with flavor oscillations included. The HL-GRBs exhibit the highest flux and the kaon contribution affects the high-energy tail of the spectra in all cases. 
\label{fig:numuosci}}  
\end{figure}

Figure~\ref{fig:numuosci} shows the expected $E^2_\nu F^{\nu_\mu}_{\rm osc}$ (computed through Eqs.~\ref{eq:siglesource} and \ref{eq:nuoscill}) as a function of the energy for a typical HL-GRB  ($\tilde{L}_{{\rm iso}} = 10^{52}\ {\rm erg}~{\rm s}^{-1}$, blue line), LL-GRB ($\tilde{L}_{{\rm iso}} = 10^{48}\ {\rm erg}~{\rm s}^{-1}$, green line), and for a sGRB ($\tilde{L}_{{\rm iso}} = 10^{51}\ {\rm erg}~{\rm s}^{-1}$, red line) located at $z = 1$ and including contributions from pion, kaon and muon decays. 
Note as the HL-GRBs give the highest flux as expected, while the LL-GRB
and the sGRB neutrino spectra are two-three orders of magnitude smaller
than the HL-GRB one for the adopted input parameters. 
 Note as by adopting the analytical prescription developed in Sec.~\ref{sec:cooling}, we
find that our estimation of the neutrino flux from GRBs gave results close to the ones obtained adopting 
numerical routines in~\cite{Mucke:1999yb,Hummer:2011ms,Aartsen:2014aqy}, by adopting their same GRB inputs.

\newpage
\section{High-energy diffuse neutrino background from gamma-ray bursts}\label{sec:nuback}
 In this section, we present our results on the high-energy diffuse neutrino background from GRB fireballs.
We first discuss the expected neutrino background within the {\emph{canonical}} model in terms of the astrophysical 
uncertainties on the local GRB rates and luminosity functions (see Table~\ref{table:LF}), then 
we study the dependence of the high-energy diffuse neutrino flux from the model parameters for each GRB family (see Table~\ref{table:model}).

\subsection{Expected diffuse background and uncertainties on the local rate and
  luminosity function of each GRB family}
\label{sec:nubackastro}
The diffuse neutrino  intensity from each GRB component (X) can be defined in
terms of the gamma-ray luminosity function, through $\Phi_{\rm
X}(\tilde{L}_{\rm iso}) d\tilde{L}_{\rm iso} = \Phi_{\rm 
X}(\tilde{L}_{\nu}) d\tilde{L}_{\nu}$ with $\Phi$ the LF introduced in
Sec.~\ref{sec:observations} (normalized to unity after integration over
luminosity):
\begin{eqnarray}
I_{{\rm X}}(E_\nu) = \int_{z_{\rm min}}^{z_{\rm max}} dz \int_{\tilde{L}_{\rm min}}^{\tilde{L}_{\rm max}} d\tilde{L}_{\rm iso} \frac{c}{4 \pi H_0 \Gamma} \frac{1}{\sqrt{\Omega_M (1+z)^3 + \Omega_\Lambda}} R_{\rm X}(z) \Phi_{\rm X}(\tilde{L}_{\rm iso}) \left(\frac{dN_{\nu_\mu}}{dE^{\prime}_\nu}\right)_{\rm osc}\ . 
\end{eqnarray}
In the numerical computation of the neutrino background, we assume  $z_{\rm min} =  0$ and $z_{\rm max} = 11$, $\tilde{L}_{\rm iso} \in [\tilde{L}_{\rm min}, \tilde{L}_{\rm max}]$ with $\tilde{L}_{\rm min}$ and $\tilde{L}_{\rm max}$ defined as in Table~\ref{table:LF} for each family X, and $E'_\nu = E_\nu (1+z)/\Gamma$. 
Note as  the chosen values  for $t_v$ and $\Gamma$  (Table~\ref{table:LF}) should guarantee us to extrapolate an average description of the whole  GRB population. 
  However,  our estimation of the diffuse neutrino emission from GRBs also depends on  parameters such as $\epsilon_e$, $\epsilon_B$, $\Gamma$ and $h_{\gamma p}$ that are currently poorly constrained from observations  (see discussion in Sec.~\ref{sec:parameters}) and should therefore be considered with caution.

For each population X,  we implement the analytical recipe described in Sec.~\ref{sec:cooling}  and automatically define the neutrino energy spectrum  according to the specific hierarchy among the different cooling processes for each   ($\tilde{L}_{\rm iso}, z$). Note as for  luminosities and redshifts different than the ones adopted in Figs.~\ref{fig:timespiHL}, \ref{fig:timespiLL} and \ref{fig:timespis}, the hierarchy among the cooling times changes. For example,  we find that the adiabatic cooling becomes  relevant for pions and kaons  when $\tilde{L}_{\rm iso}$ is on the lower tail of the studied luminosity interval for all the three GRB families.

We do not include  HL-GRBs and sGRBs whose parameters ($\tilde{L}_{\rm
iso}, z$) violate the condition  $\tau_{\gamma\gamma} \le 1$
(Eq.~\ref{eq:taugammagamma}) in our calculations.  However, for the
assumed input parameters, $\tau_{\gamma\gamma} > 1$ is
realized only for sources with $z > 7$ and with luminosities at the
upper extreme of their interval. Therefore, our computation might
underestimate the expected diffuse flux only by a few
$\%$ since the diffuse neutrino flux is not affected
from sources at $z > 7$.

\begin{figure}
\begin{center}  
\includegraphics[width=0.9\columnwidth]{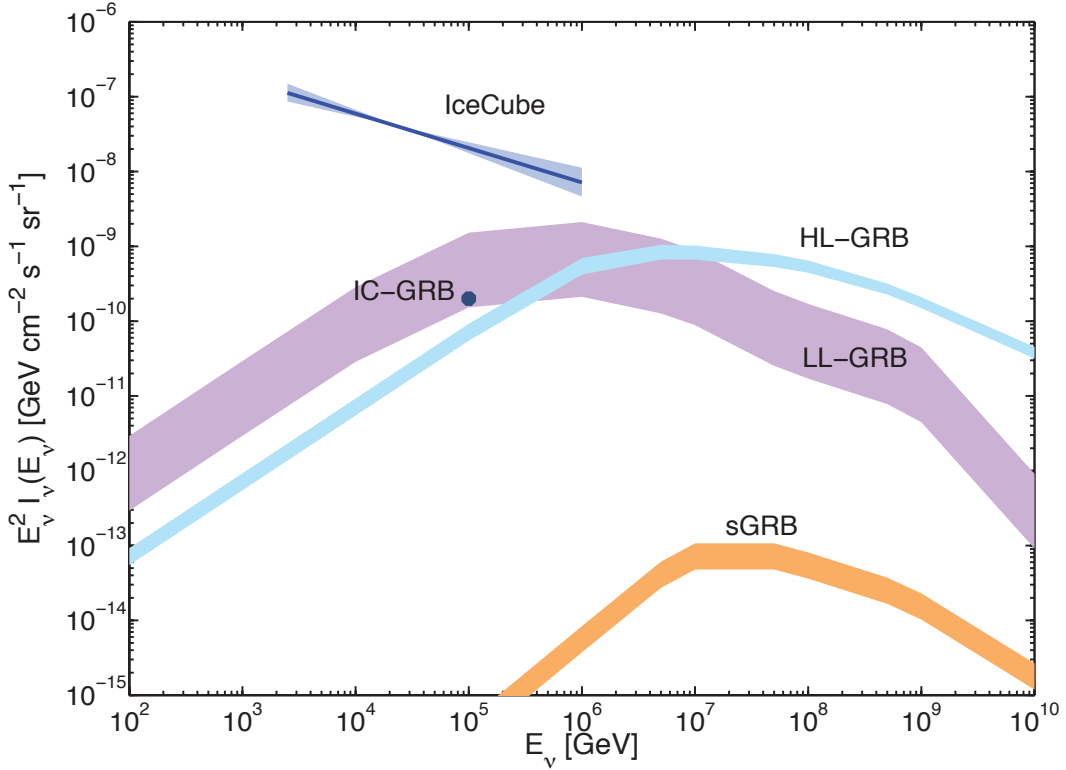}  
\end{center}  
\caption{Diffuse $\nu_\mu$ intensity as a function of the neutrino
 energy after flavor oscillations for the HL-GRB (blue band), LL-GRB
 (violet band) and sGRB (orange band) families. The
 bands represent uncertainties related to the luminosity functions and local rates
 (Table~\ref{table:LF}), whereas all the other GRB parameters are fixed
 to the \emph{canonical} values. The best fit estimation of the high-energy
diffuse neutrino flux as in~\cite{Aartsen:2014muf} is plotted in light
 blue, while the blue dot (IC-GRB) marks the upper
 limit of the GRB diffuse neutrino flux from the IceCube
 Collaboration~\cite{Aartsen:2014aqy}.  The diffuse neutrino background
 from GRB fireballs is smaller than  the observed
 high-energy IceCube neutrino flux  in the sub-PeV energy range and it scales differently as a function of the neutrino energy. \label{fig:diffuse}}
\end{figure}

Figure~\ref{fig:diffuse} shows the diffuse high-energy neutrino
intensity for the HL-GRB (light-blue band), LL-GRB (violet band) and
sGRB (orange band) components as a function of the neutrino energy. Each
band takes into account  the uncertainty due to the LF determination as
from  Table~\ref{table:LF}.

Our results should be compared with the recent IceCube discovery of
high-energy neutrinos~\cite{Aartsen:2014muf} (blue line in
Fig.~\ref{fig:diffuse}), whose sources are still unknown as well as with
the unsuccessfull searches on GRBs from the IceCube
telescope~\cite{Abbasi:2009ig,Abbasi:2011qc,Abbasi:2012zw,Aartsen:2013dla,Aartsen:2014aqy}. 
The total estimated diffuse flux from the GRB prompt emission can be as large
as    $2 \times 10^{-9}$~GeV~cm$^{-2}$~s$^{-1}$~sr$^{-1}$ at $10^6$~GeV
and it is therefore slightly lower than the best fit  of the high-energy neutrino
flux detected with IceCube (blue line in Fig.~\ref{fig:diffuse}) between
$25$~TeV and $1.4$~PeV.  Our estimated total diffuse emission is smaller than the IceCube flux at lower energies 
and it scales differently as a function of the neutrino energy. 
This implies that, assuming that the fireball model properly describes
the GRB neutrino emission, GRBs cannot be the major contributors to the
observed IceCube flux for the sub-PeV region.

IceCube results based on the monitoring of 505 observed GRBs are
presented in~\cite{Aartsen:2014aqy} with four years of data.
Extrapolating from the high-energy neutrino flux recently detected, it
is estimated in~\cite{Aartsen:2014aqy} that the GRB diffuse emission
at $100$ TeV should be smaller than $2 \times
10^{-10}$~GeV~cm$^{-2}$~s$^{-1}$~sr$^{-1}$ (blue dot in
Fig.~\ref{fig:diffuse}, IC-GRB).  Although  it might   be that such
extrapolation on the GRB emission underestimates the real GRB diffuse
flux, as IceCube is sensitive to the first break
energy that occurs for $E_\nu > 100$~TeV (see Fig.~\ref{fig:numuosci}), 
such  upper bound is very close  to our results for the  HL-GRB family. In fact our estimated high-energy neutrino intensity at $E_\nu =
100$~TeV is   $8 \times 10^{-11}$~GeV~cm$^{-2}$~s$^{-1}$~sr$^{-1}$ for HL-GRBs.
Comparing  the contribution from the HL-GRB component with the exclusion limits presented in Fig.~1
of~\cite{Aartsen:2014aqy}, we conclude that our results are compatible
with current observations  and in rough agreement with the updated 
computation of the  Waxman-Bahcall flux~\cite{Waxman:1997ti} proposed by the IceCube Collaboration (see Fig.~1 of Ref.~\cite{Aartsen:2014aqy}). 
 Rescaling the HL-GRB flux in Fig.~\ref{fig:numuosci} to the proton-to-photon luminosity ratio 
 adopted by the IceCube Collaboration ($h_{p\gamma}=0.1$), 
at $5 \times 10^6$~GeV, our predicted stacking flux for 505 sources is $9 \times
10^{-3}$~GeV~cm$^{-2}$, which is roughly in agreement with the prediction
presented in Fig.~2 of~\cite{Aartsen:2014aqy} for the standard GRB
model.\footnote{Note that we  focus on the HL-GRB family  in the comparison with the IceCube results on stacking searches 
as those are the sources for which there is higher probability to have a photon-neutrino correlation adopted in the IceCube procedure
for the GRB discrimination.}

The total expected neutrino emission from GRBs might be
further enhanced by  GRBs that do not trigger the detector because of
their low flux or because they are dark in
gamma-rays~\cite{Liu:2011cua,Ando:2005xi,Razzaque:2004yv}. However,
such contribution is  not discussed in this work. We also estimated
that the contribution from $pp$ interactions should be subleading
with respect to the one from $p\gamma$ interactions   in the energy range of relevance. 
We expect that the overall normalization of the
expected neutrino background from GRBs should be independent from the
employed GRB model for fixed  GRB parameters (see, e.g.~Fig.~1 of
\cite{Zhang:2012qy}), although peaking in slightly different energy
intervals.

\subsection{Expected diffuse background and uncertainties on the jet parameters}
\label{sec:parameters}
\begin{table}
  \begin{center}
  \caption{Variability range of the GRB model parameters adopted in the estimation of the diffuse high-energy neutrino flux.
  The variability time $t_v$ is expressed in s.}
  \vspace{2mm}
  \begin{tabular}{rrrrrrr}\hline   \hline
   \  & $\Gamma_{\rm min}$ & $\Gamma$ & $\Gamma_{\rm max}$  & $t_{v,{\rm min}}$ & $t_{v}$ & $t_{v,{\rm max}}$ \\ 
   \hline 
   HL-GRB & $100$ & $500$ & $1000$ & $10^{-3}$ & $0.1$ & $1$\\
 LL-GRB & $2$ & $5$ & $20$ & $10$ & $100$ & $200$\\
   sGRB & $100$ & $650$ & $1000$ & $10^{-3}$ & $10^{-2}$ & $0.05$\\
   \hline \hline \label{table:model}
  \end{tabular}
 \end{center}
\end{table}

Up to now, we relied on the \emph{canonical} models for each GRB family. In this Section, we adopt the GRB luminosity functions and local rates corresponding to the upper limits of the expected diffuse backgrounds plotted in Fig.~\ref{fig:diffuse} 
and discuss how our estimation depends on  the photon-to-proton luminosity ratio $h_{\gamma p}$, the bulk Lorenz factor $\Gamma$ and the variability time $t_v$.

\begin{figure}
\begin{center}  
\includegraphics[width=0.7\columnwidth]{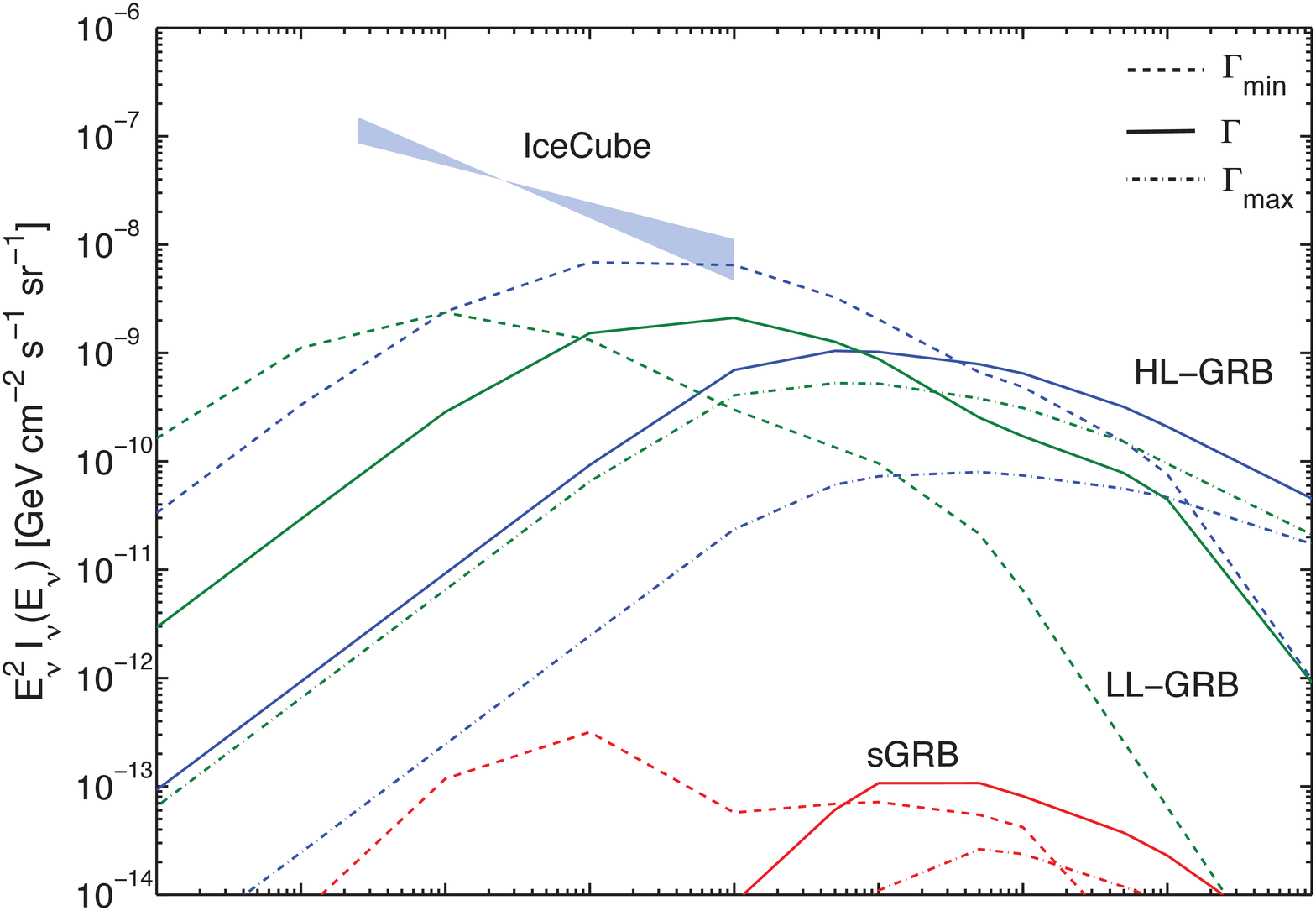}\\  
\hspace{2mm}\includegraphics[width=0.71\columnwidth]{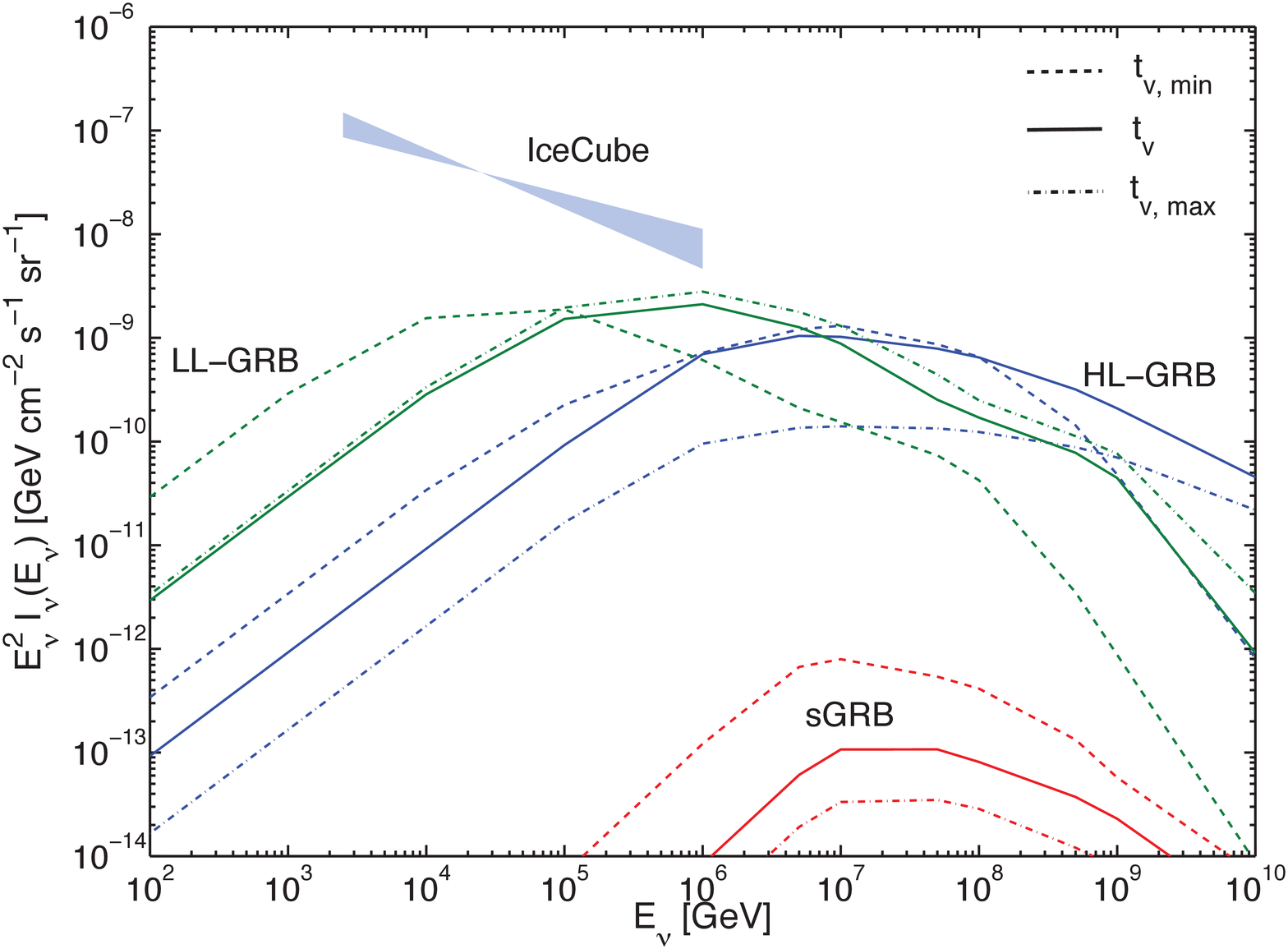}  
\end{center}  
\caption{Diffuse $\nu_\mu$ intensity as a function of the neutrino energy after flavor oscillations for the HL-GRB (blue), LL-GRB (green) and sGRB (red) families, for 
different values of $\Gamma$ (top panel) and $t_v$ (bottom panel) as from  Table~\ref{table:model}.  
The best fit estimation of the high-energy
diffuse neutrino flux as in~\cite{Aartsen:2014muf} is plotted in light
 blue.  \label{fig:diffusegammatv}} 
\end{figure}

\begin{itemize}
\item[-] \emph{Dependence on the photon-to-proton luminosity ratio $h_{\gamma p}$.} In our canonical model we assumed $h_{\gamma p} = 10^{-2}$
for the HL-GRB and the sGRB components. Such a choice is consistent with a coherent picture of neutrinos, gamma-rays and cosmic rays assuming that GRBs
are main sources of the observed UHECR  flux (see, e.g.,~\cite{Baerwald:2014zga,Asano:2014nba,Hummer:2011ms,Ahlers:2011jj} 
for  dedicated discussions). Current data suggest lower bounds for this parameter, but its precise value has not yet been fixed. Variations of $h_{\gamma p}$ correspond to an energy-independent scaling of the neutrino flux shown in Fig.~\ref{fig:diffuse} as from Eq.~(\ref{eq:normpi}).   

Assuming that the cosmic ray energy budget of the HL-GRB and LL-GRB populations are comparable, $h_{\gamma p}$ for the LL-GRB component should
 be roughly three orders of magnitude smaller than the HL-GRB one (because of the difference in  $\tilde{L}_{\rm iso}$ of the two populations). 
However, $h_{\gamma p} = 10^{-5}$ would give a diffuse neutrino flux above the current IceCube observed flux. We therefore chose our canonical value in such a 
way to boost the neutrino emission with respect to the photon one  without violating the IceCube current  bounds~\cite{Aartsen:2014muf}. We stress, however, that this
	 parameter is currently unconstrained and scalings of the  LL-GRB intensity with respect to the one presented here are not excluded yet. See also discussions in Refs.~\cite{Dermer:2014vaa,Murase:2006mm,Nakar:2015tma,Murase:2013ffa}.

\item[-] \emph{Dependence on the bulk Lorenz factor $\Gamma$.} As
	 mentioned in Sec.~\ref{sec:spectra},  we
	 fixed $t_v$ of each GRB family to the observed values in the \emph{canonical} model and determined  $\Gamma$ to guarantee optically thin GRBs for the whole $(\tilde{L}_{\rm iso},z)$ parameter space. Here, we loosen the
$\tau_{\gamma\gamma} \le 1$ constraint and study how the diffuse emission of each GRB family varies as a function of $\Gamma$.  For the HL-GRB and sGRB components, we define $\Gamma_{\rm min}$ and $\Gamma_{\rm max}$ inspired from {\it{Fermi}} data~\cite{Bregeon:2011bu,Cenko:2010cg,Dermer:2014vaa}, as reported in  Table~\ref{table:model}. The $\Gamma$ factor of LL-GRBs is poorly constrained due to the scarce statistics collected  on these sources up to now. However,  $\Gamma$ of a few seems to be favored (see, e.g.,~\cite{Toma:2006iu,Dermer:2014vaa} and references therein). We therefore consider  $\Gamma_{\rm min} = 2$ and $\Gamma_{\rm max} =  20$. The adopted intervals in $\Gamma$ for each GRB family are summarised in Table~\ref{table:model}. 

Figure~\ref{fig:diffusegammatv} (top panel) shows the correspondent diffuse GRB intensity as a function of the $\Gamma$ parameter. In general, 
the expected diffuse neutrino intensity increases as $\Gamma$ decreases. Besides the intensity normalization, $\Gamma$ also affects the neutrino break energies
as from Sec.~\ref{sec:cooling}. Note that, by adopting a wide range of variability for the HL-GRB component, the diffuse intensity could even be comparable
with the IceCube high-energy neutrino flux. However, an average $\Gamma=100$ for the HL-GRB population does not guarantee optically thin sources for any $\tilde{L}_{\rm iso}$ and $z$, 
besides being disfavored from the most recent IceCube results~\cite{Aartsen:2014aqy}. 

\item[-] \emph{Dependence on the variability time $t_v$.} The variability time $t_v$ has been fixed in the 
\emph{canonical} model to the average value preferred from observations~\cite{MacLachlan:2012cd,Dermer:2014vaa,Liu:2011cua}. Here, we study how the diffuse intensity  
changes for a minimum variability time $t_{v,{\rm min}}$ and a maximum
	 one ($t_{v,{\rm max}}$) defined in Table~\ref{table:model} as suggested from recent data~\cite{MacLachlan:2012cd,Dermer:2014vaa,Liu:2011cua}. Note that we consider  $t_{v,{\rm min}}$ one order of magnitude smaller than the one
defined for example in Fig.~4 of Ref.~\cite{MacLachlan:2012cd} assuming that current experimental sensitivity might not allow to detect even smaller variability times. 

Figure~\ref{fig:diffusegammatv} (bottom panel) shows the correspondent diffuse GRB intensity as a function of the variability time $t_v$.
Similarly to $\Gamma$, the variability time affects the normalization and the neutrino break energies as from Sec.~\ref{sec:cooling}. \end{itemize}

In all the studied scenarios, GRBs cannot explain the total detected high-energy IceCube neutrino flux in the sub-PeV energy range. However, for certain choices  of the 
model parameters, the diffuse neutrino emission from GRBs  can reach the IceCube band around PeV energies.
 
\section{Discussion and conclusions}\label{sec:conclusions}

Gamma-ray bursts (GRBs) are considered to be high-energy neutrino
emitters and candidate sources of the ultra-high energy cosmic rays. We
elaborate an analytical model to compute the neutrino flux from GRBs
based on the fireball picture, by including the pion, kaon and muon
decay contributions and taking into account radiative, adiabatic and
hadronic cooling processes. 

We revisited the fireball model for the GRB neutrino
emission~\cite{Waxman:1997ti,Hummer:2011ms}, providing a simple
analytical recipe to compute the GRB  neutrino flux. For the first time,
we included the kaon and muon contribution other than the pion one and
discussed the role of adiabatic, hadronic and radiative coolings.   Moreover, for the first
time, we investigated the relevant cooling processes  for different GRB
families, i.e., long- (divided in low-luminosity and high-luminosity)
and short-duration bursts and conclude that while the hadronic cooling
is negligible for the HL-GRB and  sGRB families, the radiative cooling is
always relevant, and the adiabatic one is not negligible for muons, and for 
pions of LL-GRBs and
sGRBs. 

By adopting up-to-date luminosity functions corrected for beaming
effects, we extrapolated the expected diffuse neutrino flux from the
long- (divided in low- and high-luminosity) and short duration
bursts. Assuming that each burst has typical parameters inferred from
observations and luminosity varying with the redshift as prescribed from
the luminosity functions in Sec.~\ref{sec:observations}, we found that
the estimated diffuse background intensity could be as large as
 $2 \times 10^{-9}$~GeV~cm$^{-2}$~s$^{-1}$~sr$^{-1}$. The low-luminosity
GRBs, being the most abundant ones, seem to dominate the overall diffuse intensity in the sub-PeV region, 
while high-luminosity GRBs are the major source at larger energies. Such conclusions are also supported 
from variations of the model parameters within the range allowed from observations.
 We conclude that GRBs do not appear to be  the leading sources originating the 
observed IceCube neutrino flux  in the sub-PeV region, if the latter is interpreted in terms of
unresolved sources.  Very recently, the possibility that the PeV
neutrinos originate by GRB cosmic rays interacting with the interstellar
gas while propagating in the host galaxy has also been
investigated~\cite{Wang:2015xpa}. However, also such an estimation
appears to predict an insufficient neutrino flux.

Previous estimates of the diffuse neutrino emission from the
high-luminosity GRBs~\cite{Gupta:2006jm,He:2012tq,Cholis:2012kq} found a
total diffuse flux from GRBs of the order of $10^{-8}~\mathrm{GeV}\
\mathrm{cm}^{-2}\ \mathrm{s}^{-1}$~sr$^{-1}$ depending on the
luminosity function and redshift evolution adopted. Our results are in
agreement with the more conservative ones presented in
Ref.~\cite{Liu:2012pf},  where an analysis on triggered and un-triggered
sources has been conducted, normalizing the triggered GRBs through the
ones detected with {\it Fermi}/GBM.
 In agreement with Refs.~\cite{Gupta:2006jm,Murase:2006mm}, we conclude  that
the neutrino flux from the low-luminosity GRBs could be comparable to the
one from the high-luminosity GRBs  and such GRBs could be the main contributors to 
the IceCube high-energy neutrino flux around $E_\nu \sim$ PeV.

Our results, although  should be considered with caution given the
uncertainties on the GRB models for the neutrino emission and their 
parameters, suggest that larger exposure is required to
discriminate neutrinos from  high-luminosity GRBs in forthcoming stacking
searches. Moreover, we find that the GRBs do not appear to be  the main
sources of the sub-PeV neutrino flux observed by IceCube. 
However, if the high-energy diffuse neutrino flux results from the
superposition of several unresolved sources, for example comparing the
contribution from starburst galaxies (see Fig.~4 of
\cite{Tamborra:2014xia}) with the one from  high-luminosity and low luminosity GRBs, it appears that while
starbursts could explain the low-energy tail of
the IceCube flux up to $0.5$~PeV, the GRBs could be responsible for the high-energy tail
of the neutrino spectrum, if this has a cutoff at energies larger than a
few PeV.
Such a hypothesis could be eventually tested in the coming
years in light of the increasing IceCube statistics.

\section*{Acknowledgments} 
We are grateful to Thomas Janka, Kohta Murase, Elisa Resconi, Mike Richman, Eli
Waxman, and Ralph Wijers for useful discussions. 
This work was supported by the Netherlands Organization for Scientific
Research (NWO) through a Vidi grant.


\end{document}